\documentclass[aps,prd,twocolumn,showpacs,superscriptaddress,groupedaddress,nofootinbib]{revtex4}  
\usepackage{graphicx}  
\usepackage{dcolumn}   
\usepackage{placeins}
\usepackage{tikz}
\usepackage{soul}

\usepackage{graphicx}
\usepackage{amsmath}
\usepackage{amsfonts}
\usepackage{epsfig}
\usepackage{slashed}
\usepackage{braket}
\usepackage{subfig}
\usepackage{comment}
\usepackage{cases}
\usepackage{appendix}
\usepackage{color}
\usepackage{booktabs}

\def\bec{\begin{center}}
\def\eec{\end{center}}
\def\beq{\begin{equation}}
\def\eeq{\end{equation}}
\def\bea{\begin{eqnarray}}
\def\eea{\end{eqnarray}}

\newcommand{\sm}{\sigma^-}
\newcommand{\spl}{\sigma^+}

\begin{document}
\title{Quantum Simulation of the $N$-flavor Gross-Neveu Model}

\author{Muhammad Asaduzzaman}
\affiliation{Department of Physics, Syracuse University, Syracuse, NY 13244, USA }
\author{Goksu Can Toga}
\affiliation{Department of Physics, Syracuse University, Syracuse, NY 13244, USA }
\author{Simon Catterall}
\affiliation{Department of Physics, Syracuse University, Syracuse, NY 13244, USA }
\author{Yannick Meurice}
\affiliation{Department of Physics and Astronomy, University of Iowa, Iowa City, IA 52242, USA}
\author{Ryo Sakai}
\affiliation{Department of Physics, Syracuse University, Syracuse, NY 13244, USA }

\date{\today}

\begin{abstract}
    We discuss the use of quantum simulation to study an $N$-flavor theory of interacting relativistic
    fermions in
    (1+1) dimensions on noisy intermediate-scale quantum (NISQ) era machines.
    The case of two flavors is particularly interesting as it can be mapped to the Hubbard model.
    We derive the appropriate qubit Hamiltonians and associated
    quantum circuits. We compare classical
    simulation and density matrix renormalization group (DMRG) / time-evolving blocked decimation (TEBD) calculations
    with the results of quantum simulation on various platforms for $N=2$ and $4$-flavors. We demonstrate that the four steps needed for calculations of real-time scattering can be implemented using current NISQ devices. 
\end{abstract}

\maketitle

\section{\label{intro}Introduction}
Doing ab-initio lattice QCD calculations in real-time or at finite density 
would have a significant impact on our interpretation of hadron collider data and our understanding of nuclear matter. However, because of sign problems, such calculations  
cannot be handled efficiently with importance sampling (Monte Carlo) methods.
In contrast, these calculations could be handled 
efficiently
by using quantum devices capable
of manipulating large enough Hilbert spaces that can be mapped into those relevant for the QCD Hamiltonian.
The possibility of using universal quantum computers \cite{Jordan:2011ci,Jordan_2012,lloyd1996universal,Mezzacapo:2015bra,cervera2018exact,Yeter-Aydeniz:2018mix,Raychowdhury:2018osk,Lamm:2018siq,macridin2018digital,Klco:2018kyo,Bauer:2019qxa,Lamm:2019uyc,Lamm:2019bik,Gustafson:2019mpk,gustafson2021indexed,Gustafson:2021imb,Kharzeev:2020kgc,Honda:2021aum,
Bhattacharya:2020gpm,Ji:2022qvr}, or analog quantum simulations with cold atoms \cite{bloch2012quantum,lewenstein2007ultracold,cirac2010cold,Kapit:2010qu,Kuno:2016ipi,Martinez:2016yna,Danshita:2016xbo,Zhang:2018ufj,Davoudi:2021ney,Davoudi:2019bhy,Monroe:2019asq,Gonzalez-Cuadra:2017lvz,Nguyen:2021hyk,Aidelsburger:2021mia,Schweizer:2019lwx} has motivated 
road maps \cite{Bauer:2022hpo,Banuls:2019bmf,Kasper:2020akk,Dalmonte:2016alw,meurice2022tensor} to implement sequences of models of increasing complexity and dimension using the rapidly evolving NISQ technology \cite{RevModPhys.51.659}.

In this context, the Schwinger model has often been the first model to try \cite{Hauke:2013jga,Kuhn:2014rha,Martinez:2016yna,Klco:2018kyo,Thompson:2021eze,Nguyen:2021hyk,Shaw:2020udc,Kharzeev:2020kgc}.
However, the Gross-Neveu (GN) model with $N$ species of fermions in 1+1 dimensions is also a particularly important step in the study of relativistic fermions. 
As for QCD, this model is asymptotically free and capable of
dynamical mass generation and has a rich phase structure at finite temperature and finite density \cite{Gross:1974jv}. An efficient initial state preparation for the massive GN model in one spatial dimension has been 
developed by Moosavian and Jordan \cite{HamedMoosavian:2017koz}. Because of the limited entanglement entropy in one spatial dimension, this model can also be handled efficiently with classical computers using the density matrix renormalization group (DMRG) and the time-evolving block decimation (TEBD) methods based on matrix product states (MPS)~\cite{Moosavian:2019rxg,roose_lattice_2021,Roose:2021pba} in order to explore and validate quantum
simulations.

In this article, we show how to map 
the continuum Hamiltonian for the GN model to a spatial lattice qubit system using a Jordan-Wigner transformation~\cite{Jordan:1928wi,dargis_fermionization_1998} and derive
quantum circuits which can be used for its time evolution with first order Trotter approximation and to find the ground state
wavefunction for the massless model for a range of values of the four fermion coupling using the variational quantum eigensolver (VQE) algorithm. We then
compare Trotterized evolution of wavefunction on two platforms---the IBM-Q Guadalupe and Honeywell Quantinuum machines---with the results of exact diagonalization and DMRG/TEBD calculations.
We demonstrate that the four steps of calculations needed for real-time scattering as outlined by Jordan, Lee and Preskill (JLP) 
\cite{Jordan:2011ci,Jordan_2012}, namely, 1) vacuum preparation, 2) excitation of 
single-particle
wavepackets, 3) 
unitary time 
evolution, and 4) measurements for the final state, can all be achieved with current NISQ technology for small systems.

The lattice formulation of the GN model exhibits connections with several different condensed matter systems including the Hubbard model \cite{reiner2016emulating,stanisic2022observing,hubbard1963electron}
and the Su-Schrieffer-Heeger model of polyacetylene ~\cite{Campbell:1981dc,Chodos:1993mf,Kuno:2018pcp, roose_lattice_2021,Roose:2021pba,PhysRevLett.42.1698}. Inhomogeneous phases are present at 
finite temperature and density \cite{Basar:2009fg} and can be studied using numerical  lattice simulations  \cite{Lenz:2020bxk}, ultra cold fermionic atoms in optical lattices \cite{Bermudez:2018eyh,Ziegler:2020zkq,Ziegler:2021yua,Tirrito:2021fbj} or with transmon qubits \cite{reiner2016emulating}. The use of configurable arrays of Rydberg atoms, proposed  for the 
the Schwinger model \cite{Surace:2019dtp,Surace:2020ycc,Notarnicola:2019wzb} or the Abelian Higgs model
\cite{Meurice:2021pvj}, could also be adapted for the real-time evolution of the GN model. Complementary to this work the GN model is also often used to test new field theoretical methods and ideas, for instance for 
mass gap calculations \cite{Verschelde:1997jx} or to search for nontrivial infrared fixed points \cite{Choi:2016sxt}.

The article is organized as follows. The lattice formulation, the Jordan-Wigner transformation and the Trotter approximation of the $N$-flavor GN model in two space-time dimensions are presented in Sec. \ref{hamiltonian}. The massless two-flavor model, its symmetries and the choice of the Trotter step are discussed in Sec. \ref{two}. The preparation of the ground state with a variational quantum eigensolver is explained in Sec. \ref{ground} and the scattering of wavepackets is discussed in Sec. \ref{packet}. The formulation of the Hamiltonian for the DMRG calculation is explained in the Appendix.~\ref{dmrg_formulation}, while construction of the different building blocks of the Trotter evolution circuit is discussed in the Appendix~\ref{gates}. Finally Trotter evolution results for the four flavor GN model is presented in Appendix.~\ref{four}.

\section{\label{hamiltonian}From continuum Hamiltonian to quantum circuit}
We start from the general form of the continuum Hamiltonian for one Dirac fermion in
one spatial dimension:
\begin{equation}
    H=\int dx\; i{\psi}^\dagger \alpha\partial_x \psi+m {\psi}^\dagger\beta\psi.
\end{equation}
To discretize we place the theory on a lattice and replace the continuum derivative by
the symmetric difference operator
$\partial\to \Delta_{n^\prime\;n}=\frac{1}{2}\left(\delta_{n^\prime\;n+1}-\delta_{n^\prime\;n-1}\right)$. In addition, we employ a staggered fermion construction in which the
original field $\psi$ is replaced by a new field $\lambda$ via the 
unitary
transformation $\psi(n)=\alpha^n\lambda(n)$ where $n$ labels the lattice site. This yields
the lattice Hamiltonian
\begin{align}
    H&=\sum_{n=1}^L \Big( \frac{i}{2}  \,{\lambda}^\dagger(n)\left[\lambda(n+1)-\lambda(n-1)\right] \nonumber \\
    &\qquad+m\left(-1\right)^n{\lambda}^\dagger(n)\beta\lambda(n) \Big).
\end{align}
The sum is over all lattice sites and depending on the boundary conditions some kinetic terms at the lattice edges need to be omitted or modified.
In the usual Euclidean path integral the staggering transformation
has the effect of reducing the spinor structure of the fermion
operator to the unit matrix and in consequence
all but one of the spinor components can be discarded. In a Hamiltonian formulation one
can only do this for the derivative but not the mass term. Instead
we allocate the 2 spinor components to even and odd lattice sites corresponding to the choice
$\alpha=\sigma^x$. Choosing $\beta=\sigma^z$ then generates the staggered mass term
$m\left(-1\right)^n \lambda^\dagger(n)\lambda(n)$.
If we denote $\chi_{\rm even}=\lambda^1$ and $\chi_{\rm odd}=\lambda^2$ this can be trivially rewritten:
\begin{align}
    H&=\frac{i}{2}\sum_{n=1}^L\chi^\dagger(n)\left[\chi(n+1)-\chi(n-1)\right]\nonumber\\
    &+m\sum_{n=1}^L\left(-1\right)^n\chi^\dagger(n)\chi(n).
\end{align}
For $N$-flavors of Dirac fermions  we can then add
a four fermion term to generate a Gross-Neveu model \cite{Gross:1974jv}. Rescaling mass $m$ and the four fermion coupling $G^2$ allows us to omit the factor in front of the kinetic term:
\begin{align}
    H^{(N)}&=\sum_{n}\Bigg[i\sum_{a=1}^N\chi^{a\dagger}(n)
    \left[\chi^a(n+1)-\chi^a(n-1)\right] \nonumber \\
    &+m\left(-1\right)^n\chi^{a\dagger}(n)\chi^a(n) + G^2
    \left(\sum_{a=1}^N\chi^{a\dagger}(n)\chi^a(n)\right)^2 \Bigg].
\end{align}
The resultant Hamiltonian has a manifest $SU(N)$ symmetry.
To simulate this system on a quantum computer we first need to rewrite the theory
in terms of Pauli matrix or qubit operators. We
use the Jordan-Wigner transformation~\cite{Jordan:1928wi,dargis_fermionization_1998}
\begin{align}
    \chi^a(n)&= \prod_{b<a}P^{\left(\sigma^b\right)}(L) \prod_a P^{\left(\sigma^a\right)}(n-1) \sigma_+^a(n),
\end{align}
where 
\begin{equation}
    P^{\left(\sigma^a\right)}(n)=\prod_{y=1}^n\sigma_3^a(y)
\end{equation}
and $\sigma_\pm=\frac{1}{2}\left(\sigma_1\pm i\sigma_2\right)$.
It is straightforward, 
to show that this representation respects the fundamental anticommutator required for
fermion operators
\begin{equation}
    [\chi^{a\dagger}(x),\chi^b(y)]_+=\delta_{xy}\delta^{ab}.
\end{equation}
We have used open boundary conditions in our work. In this representation the free massive Hamiltonian becomes
\begin{align} \label{qubit_ham}
H^{(N)}_{0,m}&=\sum_{a=1}^{N} \frac{1}{2}\bigg[ i\sum_{n=1}^{L-1} \Big(\sigma^a_+(n)\sigma^a_-(n+1)-\sigma^a_-(n)\sigma^a_+(n+1)\Big)\nonumber\\
&+m\sum_{n=1}^{L}\left(-1\right)^n\left(\sigma^a_-(n)\sigma^a_+(n)\right) \bigg]+{\rm h.c} \nonumber\\
&= \sum_{a=1}^{N} \Bigg[ \sum_{n=1}^{L-1} \Big( -\sigma^a_1(n)\sigma^a_2(n+1)+\sigma^a_2(n)\sigma^a_1(n-1)\Big) \nonumber\\
&+m \sum_{n=1}^{L}\left(-1\right)^n\left(1-\sigma_3^a\right) \Bigg]
\end{align}
while the four fermi term is
\begin{equation}
    H^{(N)}_{G}=\frac{1}{2}G^2\sum_{n=1}^{L} \,\, \sum_{a=1}^{N} \,\, \sum_{b,b>a} (I-\sigma^a_3(n))(I-\sigma^b_3(n)).
\end{equation}
Notice that one can think of flavor as another lattice dimension. That is one
can imagine that the problem maps to a ladder geometry where each leg of an $N$ leg
ladder corresponds to the spatial lattice while the vertical rungs correspond to four
fermion interactions between pairs of flavors. Alternatively we can map all $N$ legs into
a single one dimensional lattice with 
the different flavors mapping into consecutive lattice sites in such a way
that
the unit cell of the lattice is of length $N$ \cite{roose_lattice_2021}.

\begin{figure}[!ht]
    \centering
    \includegraphics[width=0.5\textwidth]{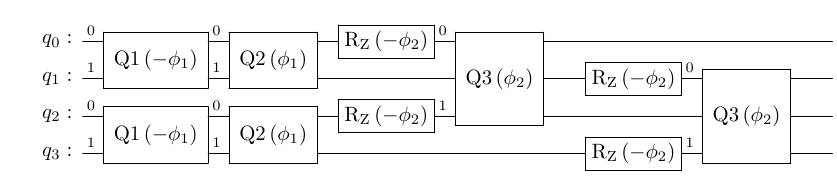}
    \caption{Schematic diagram of the quantum operations in the circuit form is shown for a single step of Trotter evolution for
    two flavors. Here, $\phi_1=2\Delta t$ and $\phi_2=G^2 \Delta t$.}
    \label{schematic}
\end{figure}

We consider Trotter evolution of the Hamiltonian with zero staggered mass. This reduces $H^{(N)}_{0,m}$ to $H^{(N)}_{0,0}$ and can be obtained from Eq.~(\ref{qubit_ham}) omitting the last term. Thus the Hamiltonian for the $N$-flavor massless staggered fermions
can be written as
\begin{equation}
    H^{(N)}_{m=0} = H^{(N)}_{0,0}+ H^{(N)}_{G}.
\end{equation}
To evolve the system in time requires exponentiation of the Hamiltonian. Since it is composed
of non-commuting pieces we have employed the first order Suzuki-Trotter approximation
for a small time step $\Delta t=t/n$ \cite{trotter1959product,suzuki1992general,suzuki1990fractal,suzuki1993improved}
\begin{equation}
    e^{-i(H^{(N)}_{0,0}+H^{(N)}_{G}) t} \approx\left(e^{-i\,H^{(N)}_{m=0} t / n} \,\, e^{-i\,H^{(N)}_{G} \frac{t}{n} }\right)^{n}+\mathcal{O}(t \Delta t), \\
\end{equation}
where the contribution of the
kinetic term $H_k$ for a fixed flavor $a$ 
can be decomposed into elementary 2-qubit operations $Q_1=\exp{\left[i \Delta t \, \sigma^a_1(n)\otimes \sigma^a_2(n+1)\right]}$ and $Q_2=\exp{\left[-i \Delta t \, \sigma^a_2(n)\otimes \sigma^a_1(n+1)\right]}$.
The four fermion interaction
term $H^{(N)}_{G}$ couples two such flavors $a$ and $b$ at the same physical site introducing additional qubit operators of the form $Q_3=\exp{\left[i\frac{\Delta t}{2} G^2 \sigma_3^a (n)\otimes \sigma_3^b (n)\right]}$ and $R_z=\exp{\left[-i\frac{\Delta t}{2} G^2 \sigma_3^a (n)\right]}$. A schematic diagram showing how these operations
are combined to generate a single time step is shown in Fig.~\ref{schematic}.
The individual gates needed to generate $Q_1$, $Q_2$, and $Q_3$ appear in 
Figs.~\ref{Q1},~\ref{Q2}, and~\ref{fig_sz_sz} in appendix~\ref{gates}, which
gives a detailed
description of how these elementary circuit blocks involving CNOT, Hadamard, and rotation
gates implement the elementary qubit operations needed for the Hamiltonian.

From this point on we will focus on the results of our quantum simulations for $N=2$ flavor GN model in the main text. 

\section{\label{two}Time evolution}

Interestingly the $2$ flavor model at $m=0$ can be mapped into the Hubbard model \cite{hubbard1963electron} at a particular
value of the chemical potential~\cite{melzer_scaling_1995}. The four fermi interaction 
clearly corresponds to a Hubbard term after identifying $n^\uparrow=\chi^{1\dagger}\chi^1$ and $n^\downarrow=\chi^{2\dagger}\chi^2$.
In addition,
the kinetic operator can be mapped
to the usual Hubbard hopping term after performing an additional
unitary transformation $\chi^a(n)\to i^n\chi(n)$.
In this case the manifest $SU(2)$ symmetry of the two flavor theory is enhanced to
$SO(4)$ which is most easily seen by 
decomposing each staggered field in terms
of real (or reduced staggered) fields via the mapping

  \begin{align}
  \left.\begin{aligned}
\chi^{1\dagger}&=\xi_1+i\xi_2 \qquad \\
\chi^1&=\xi_1-i\xi_2 \qquad \\
    \chi^{2\dagger}&=\xi_3-i\xi_4 \qquad \\
    \chi^2&=\xi_3+i\xi_4 \qquad
\end{aligned}\right\}
\end{align}

The Hamiltonian including the four fermi term can then be written
\begin{equation}
    H_{m=0}^{(2)}=\sum_n \xi^i(n)\xi^i(n+1)+\frac{G^2}{12}\epsilon_{ijkl}\xi^i(n)\xi^j(n)\xi^k(n)\xi^l(n).
\end{equation}
In this form it can be identified with recent path integral 
studies of reduced staggered fermions 
capable of symmetric mass generation in (spacetime) dimension $D\ge 2$
\cite{Ayyar:2014eua,Catterall:2015zua,Ayyar:2017qii,Ayyar:2015lrd,Butt:2018nkn}.


\begin{figure}
    \centering
    \includegraphics[width=0.45\textwidth]{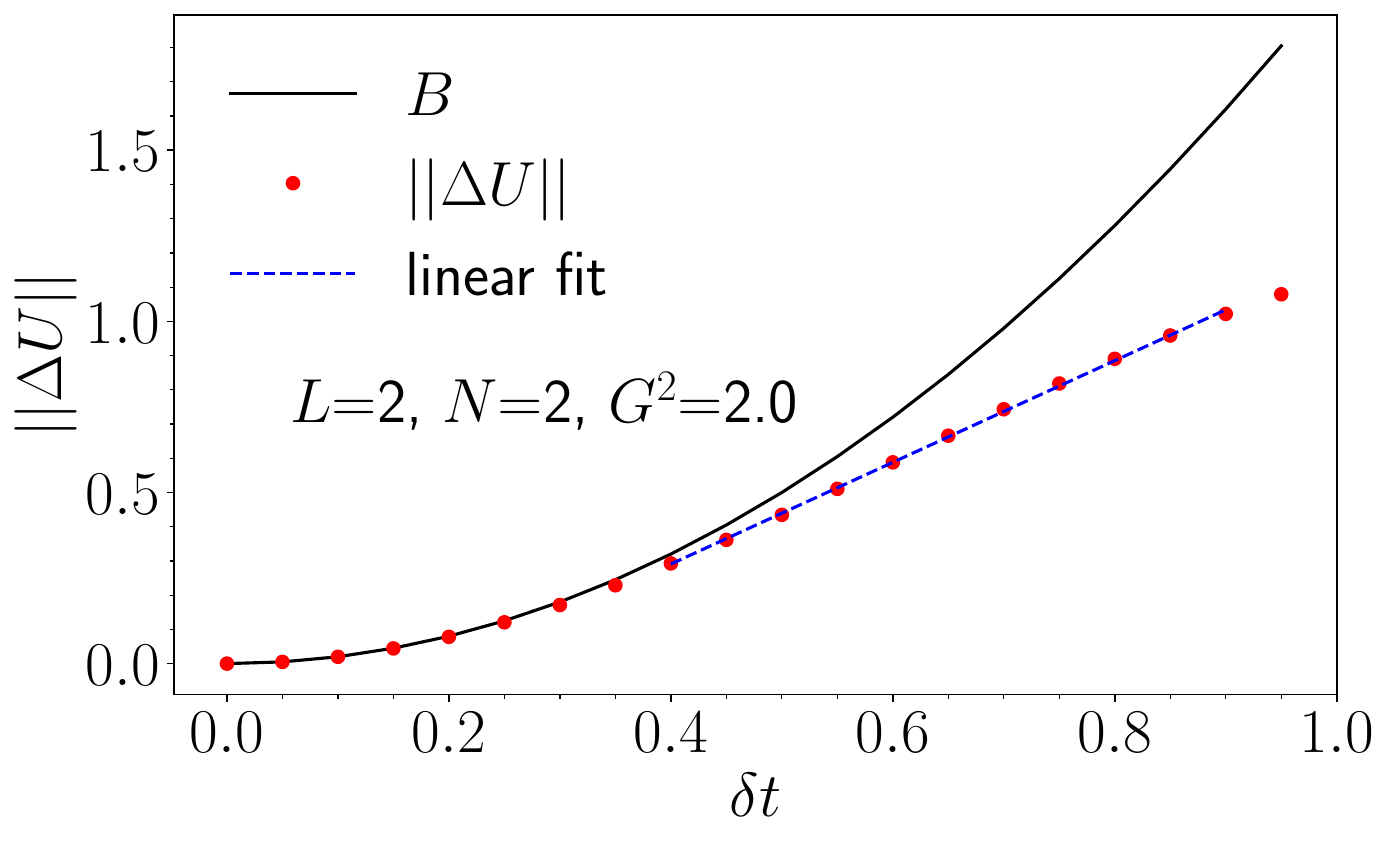}
    \caption{Comparison of the 1st order Trotter-bound ($B$) with the practical bound ($||\Delta U ||$): differences in the norm of the unitaries $||\Delta U||$ computed using matrix exponentiations of our model. }
    \label{trotter_bound}
\end{figure}

\begin{figure}[ht]
\centering
\includegraphics[width=0.48\textwidth]{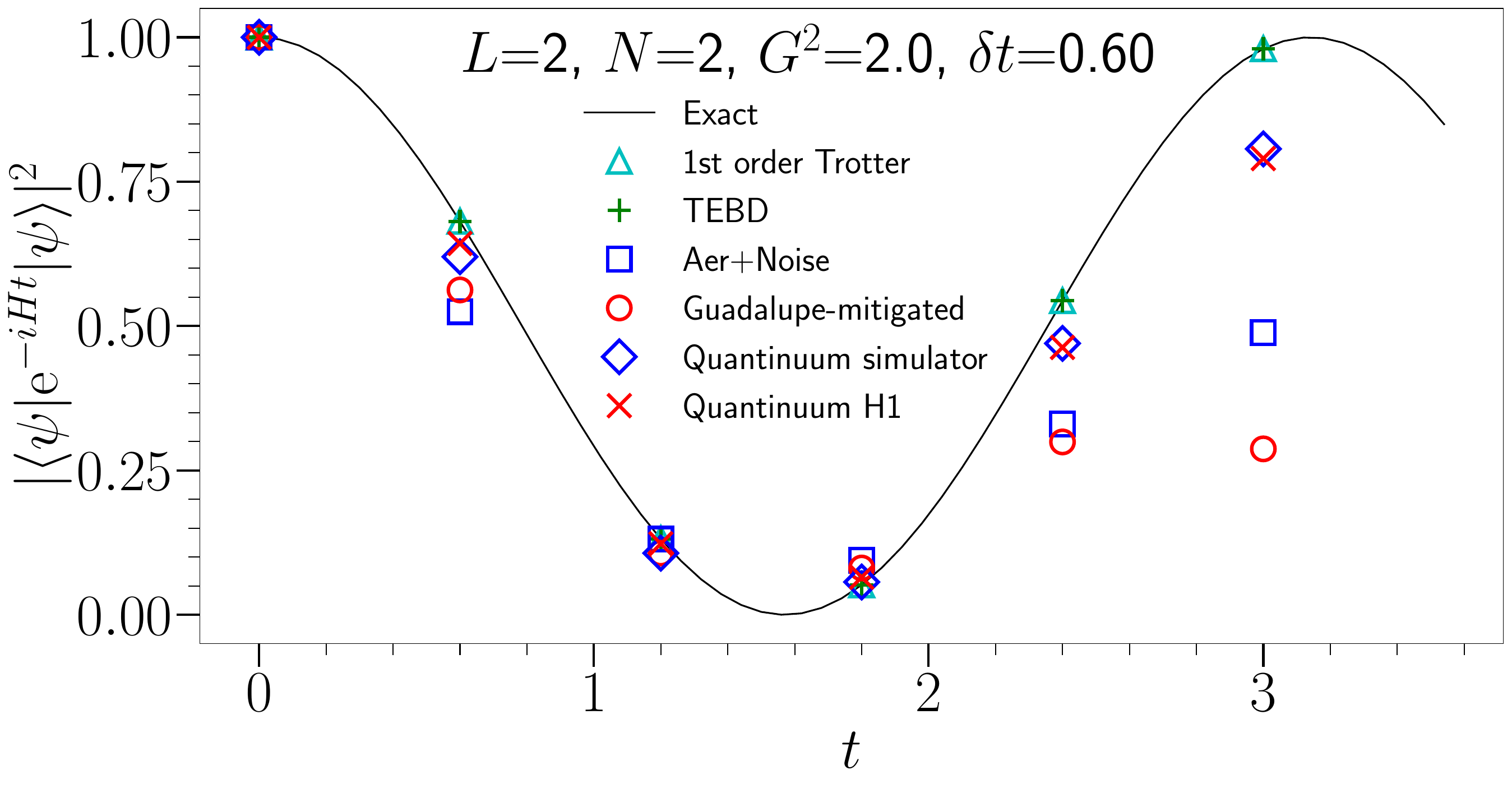}%
\caption{Trotter evolution for the $N=2$ flavor model with $L=2$ lattice sites
and $G^2=2.0, m=0.0$ and time step $\delta t=0.6$ from initial state $|\psi\rangle=|0010\rangle$. The number of shots used for Guadalupe and Quantinuum simulation are 4000 and 300 respectively.\label{Trotter1}}
\end{figure}

We have simulated the model using a first order Trotter update on both the IBM-Q Guadalupe quantum processing unit (QPU) and the
Honeywell Quantinuum platforms. The Quantinuum provider gives access to two H1 generations of QPU: H1-1 and H1-2. Results of the Trotter evolution are shown 
in Fig.~\ref{Trotter1} for $G^2=2.0$, $m=0.0$ and Trotter step $\delta t=0.6$ on a lattice with two sites. The initial wavefunction can be written in the computational basis $\ket{\psi}=\ket{0100}$. From the classical exact diagonalization analysis, it is found out that to capture the characteristics of the time-evolved wavefunction at $G^2=2.0$ we need to consider computing Trotter evolution up to a time $t\sim 3.0$. Hence, we used a large Trotter step due to practical limitations of computing Trotter evolution for large number of steps with NISQ-era machines. It has been demonstrated previously for the quantum Ising model
that the Trotter step can be taken 20 or 30 times larger compared to the theoretical bound of the first order Trotter step $\Delta t^2$ before large discretization errors are encountered \cite{gustafson2021indexed,meurice2021quantum}. We performed an identical analysis with our model and found the conclusion to be true for our model too. We compute the norm of the following operator
\begin{align}
    \Delta U &= \mathrm{e}^{-i H^{(2)}_{m=0}\delta t}-e^{-i  H^{(2)}_{0,0} \delta t} e^{-i G^2 H^{(2)}_{G} \delta t}.
\end{align}
Here the norm is defined to be the largest singular value of the corresponding operator.
The theoretical bound up to the second order in $\delta t$ of the norm in this quantity is $B= (G^{2}/2) || [ H^{(2)}_{0,0}, H^{(2)}_{G} ] || (\delta t)^{2}$. However the actual bound $||\Delta U||$ is strictly less in the region $0.55<t<0.95$, and we numerically find that the actual bound is consistent with a linear approximation in $\delta t$
\begin{equation}
    ||\Delta U|| \sim 1.49(1)\delta t -0.304(7).
\end{equation}
Figure~\ref{trotter_bound} shows the comparison of the second order Trotter bound ($B$) from 1st order Trotterization with the actual value of the difference in the norm ($||\Delta U||$). The choice of the Trotter step $\delta t=0.6$ is justified where $||\Delta U|| \sim 0.6$.
In Fig.~\ref{trotter_bound}, the quantum simulations are compared with exact diagonalization, first order Trotterization 
code, and the TEBD algorithm written 
using the ITensor library~\cite{itensor}.

Before implementing the circuit on quantum hardware
we also simulated the circuit using the device noise model. 
For the IBMQ QPU, we used the aer-simulator using a basic device noise model derived from the backend properties. The noise model incorporates a simplified model for the gate error probability of each basis gate on each qubit taking into
account the relaxation time and readout probability of each qubit. Figure~\ref{Trotter1} shows that for a small number of qubits $Q=N\times L=4$, the error 
model predicts the results from the QPU quite well out to four
Trotter steps.
We also performed analysis with the noise model of the Quantinuum machine in the native simulator of Quantinuum provider. The noise model simulator was seen to predict the Trotter evolution from the Quantinuum machine extremely accurately up to five Trotter steps.

\begin{table}[!htb]
\centering
\begin{tabular}{
|>{\centering\arraybackslash}p{2.0cm}||
    >{\centering\arraybackslash}p{1.5cm}|
    >{\centering\arraybackslash}p{1.5cm}||
    >{\centering\arraybackslash}p{1.2cm}|
    >{\centering\arraybackslash}p{1.2cm}|}
   
\multicolumn{1}{c}{} & \multicolumn{2}{c}{IBMQ} & \multicolumn{2}{c}{Quantinuum} \\ \hline
 Trotter step, $n$ & $d$  & CX    & $d$ & CX  \\ \hline
 1 & 24 & 4   & 10 & 4  \\ \hline
 2 & 47 & 18  & 22 & 12 \\ \hline
 3 & 70 & 32 &  34 & 20  \\ \hline
 4 & 93 & 46  & 46 & 28 \\ \hline
 5 & 116 & 60 & 58 & 36 \\ \hline  
\end{tabular}
\caption{Circuit depth $d$ and the number of CNOT (CX) gates required for the implementation of the Trotter evolution for different numbers of steps $n$. \label{circuit_depth}}

\end{table}

Table \ref{circuit_depth} shows the circuit 
depth of the implemented circuits~\footnote{ The original circuit is transpiled before submitting the circuit to the QPU in order to express the circuit in terms of the native gates and optimize the mapping of qubits to the QPU. The depth noted here is the depth for the transpiled circuit.}. We also note the number of CNOT gates for each case. Notice that twice as many CNOT gates are required in the case of the IBM-Q relative
to the Honeywell machine. The reduced number of gates for the latter
machine reflects the all-to-all qubit connectivity which
eliminates the need for additional SWAP operations requiring
three CNOT gates.  These facts
account for the observed difference in the two platforms.  Our overall conclusion
is that the layout of the physical qubits plays a very important
role in the efficiency with which quantum simulation can be accomplished
on NISQ era hardware.

\begin{figure*}[!htb]
    \centering
    \includegraphics[width=\textwidth]{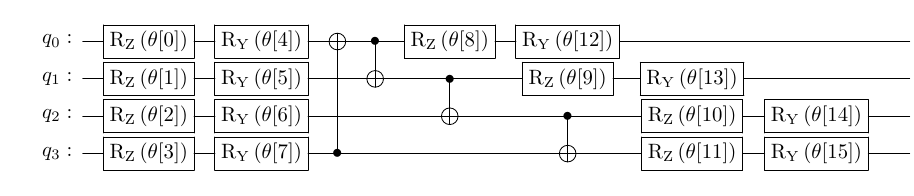}
    \caption{Quantum circuit used in estimating the ground state wavefunction for 2 flavors}
    \label{GSckt}
\end{figure*}

\section{\label{ground}Computation of the ground state}
One of the fundamental goals of investigating any interacting lattice fermionic model is to
understand its phase structure. The first step of doing that is to prepare the ground state. In our work, we designed a quantum circuit that is suitable for use with the Variational Quantum Eigensolver (VQE) algorithm \cite{mcclean2016theory} to
determine the ground state wavefunction of the system as a function of the four
fermi coupling. 
The inputs to the algorithm are the qubit Hamiltonian and a parametrized quantum
circuit whose function is to evaluate the expectation value of the Hamiltonian on a trial
ground state wavefunction. The algorithm uses a classical computer to minimize
the energy of the state with the quantum circuit being used to evaluate
the expectation value of the Hamiltonian on the trial wavefunction at each stage of
the iteration.
The wavefunction ansatz for
the ground state for two flavors
is the well-known Hardware Efficient Approximation (HEA) \cite{kandala2017hardware} shown in Fig.~\ref{GSckt}, where a set of rotation
angles $\theta_a,a=0, \, 1\ldots 15$ are used
as variational parameters. For a $Q$ qubit lattice model, the expression of the HEA wavefunction ansatz in terms of rotation gates $R$ and 2-qubit entangling operators $U^{mn}$ which entangle the m\textsuperscript{th} qubit and n\textsuperscript{th} qubit can be written as
\begin{align}
    \ket{\psi} &=\sum_{i=1}^{Q} \Bigg[  (\prod_a {R^i_a}) ( \prod_b U^{i,i+1}_b )  (\prod_c {R^i_c})  \Bigg] \ket{\psi_0}  \nonumber\\ 
    & \equiv M_\alpha( \{\theta \} )  \ket{\psi_0}.
\end{align}

Here, the subscript in $R^i_a$ denotes different rotation operators along different axes with $a,c=\{x,y,z\}$, and the subscript in $U_b$ denotes different entangling operators $U_b \in \{CX,\,CY,\,CZ,\,CH,\,CRZ,\,CU,\cdots \}$. The number of product terms in each part can be varied and a suitable number of terms can be chosen for approximating the ground state. In principle, the operator $M_\alpha (\{\theta\})$ can be repeated as many times as needed with a new set of parameters for each $M_\alpha$ block. Thus, in general, the structure of the HEA ansatz can be written as
\begin{align}
    \ket{\psi}= \prod_{\alpha( \{ \theta\} )} M_\alpha( \{\theta \} ) \ket{\psi_0}.
\end{align}

\begin{figure}
    \centering
    \includegraphics[width=0.48\textwidth]{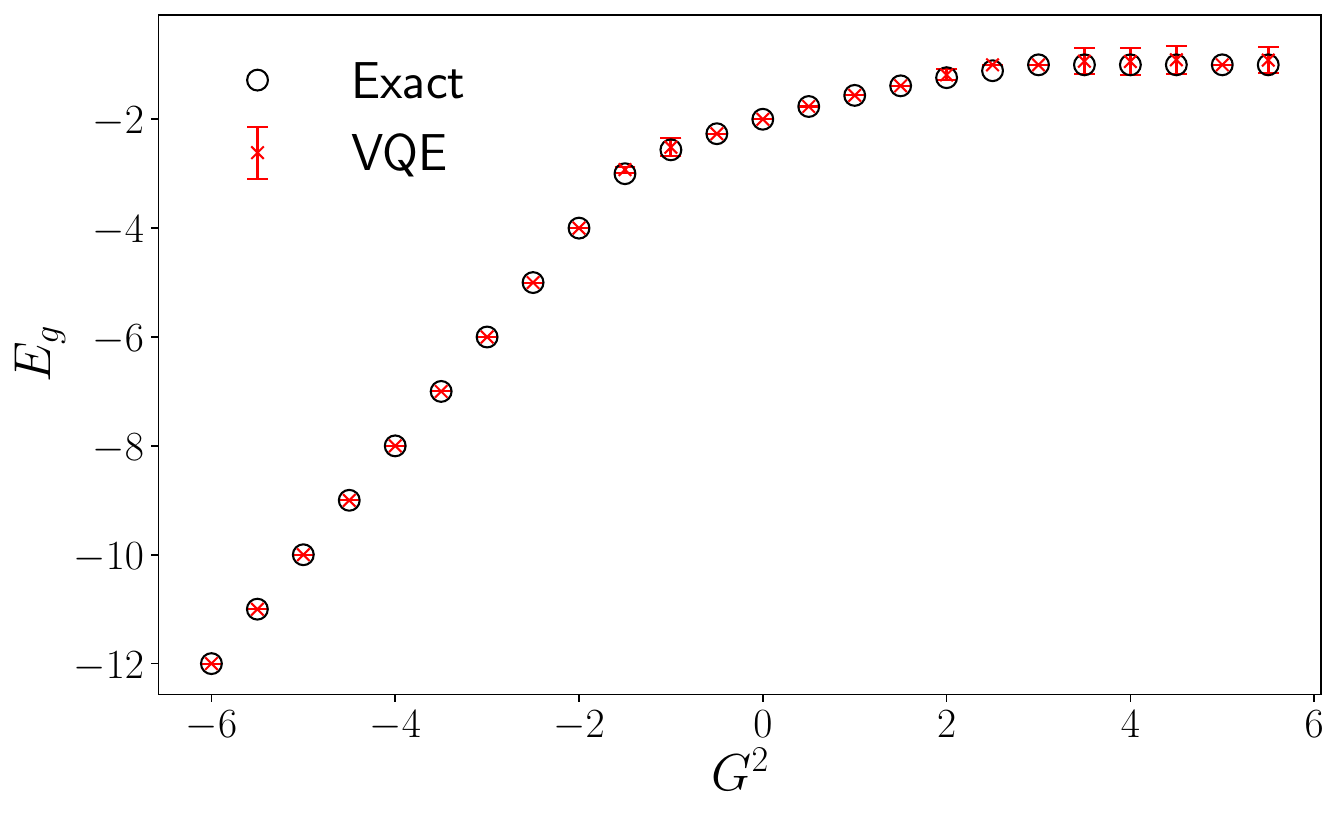}
    \caption{Ground state energy computed from the VQE compared with the results of the exact diagonalization. DMRG results are not shown here as they match exactly with the exact diagonalization result.}
    \label{GSE_comparison}
\end{figure}

\begin{figure}[!h]
    \centering
    \includegraphics[width=0.45\textwidth]{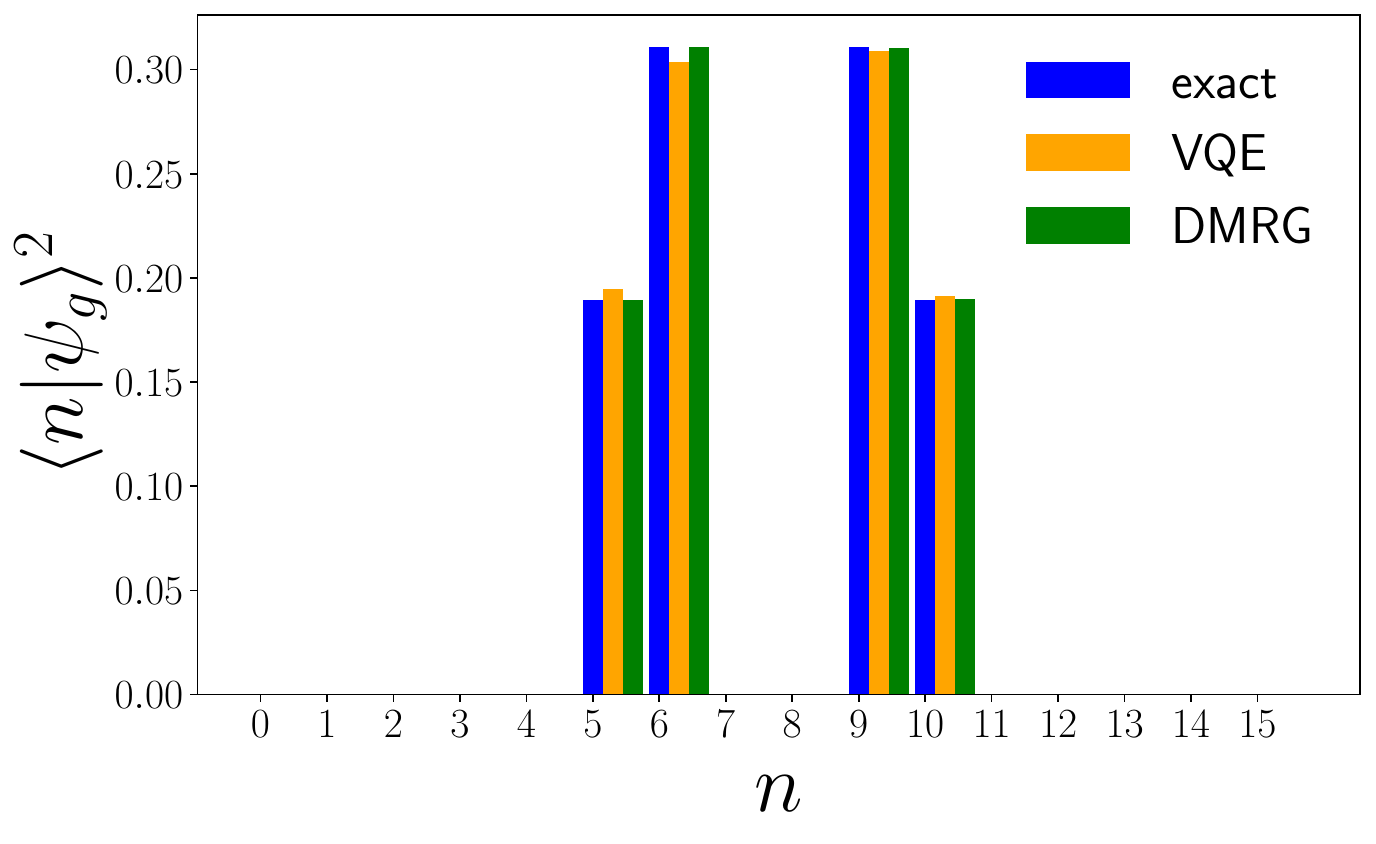}
    \caption{Projection of the ground state $\ket{\psi_g}$  on the computational basis $\ket{n}$ derived from the VQE. VQE result is compared with the exact diagonalization and results obtained from DMRG.}
    \label{GS}
\end{figure}

\begin{figure}[!h]
    \centering
    \includegraphics[width=0.45\textwidth]{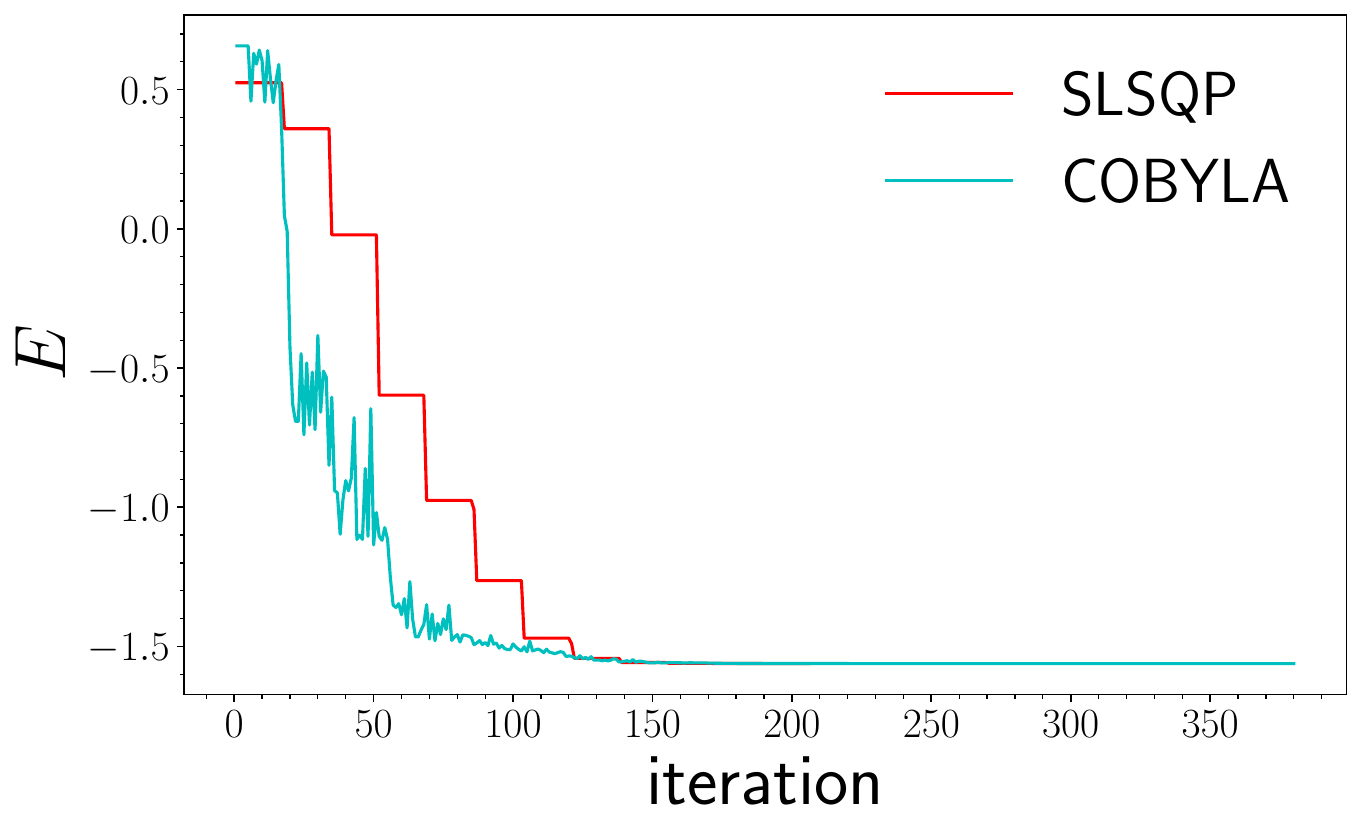}
    \caption{Energy minimization using VQE alogorithm with two different classical optimizers at $G^2=1.0$.}
    \label{findmin}
\end{figure}

Repeating the block ($M_\alpha$) $N$-times increases the number of parameters by the same
factor. The number of parameters $p$ needed is bounded by $p<QN\sum_i \ell_i$ where
$\ell_1$ and $\ell_3$ denote the number of rotation layers
in the first and the last stage respectively and $\ell_2$ the number
of layers of entangling gates.
VQE then uses Ritz's variational principle to update the parameters $\theta_i$ \cite{macdonald1933successive}.
We used the Constrained Optimization by Linear Approximation  (COBYLA) optimizer \cite{powell1994direct,powell1998direct,powell2007view} with the statevector simulator of Qiskit to  determine the change of the
parameters at each stage of the iteration. The COBYLA optimizer is based on a linear approximation of the objective functions and the constraints. To verify whether the `true' ground state is reached, we compared the results obtained from the COBYLA optimizer with the SLSQP optimizer \cite{kraft1988software}. SLSQP uses Sequential Least Squares Programming to minimize a function of several variables. Any combination of bounds, equality and inequality constraints can be incorporated in the SLSQP optimization routine. Figure~\ref{findmin} shows a typical
relaxation of the energy to the ground state at $G^2=1.0$ for both optimizers. We assumed the algorithm reaches the ground state when two successive iteration match up to the fourth order after the decimal point.
For the same $G^2$, the projection of the obtained ground state $\ket{\psi_g}$ on the computational basis $\{\ket{n}\}$ is shown and compared with results from the exact diagonalization and the DMRG result in Fig.~\ref{GS}. Similar analysis was performed at different values of $G^2$ and the computed ground state energy from the Variational Quantum Eigensolver is compared with the exact diagonalized result in Fig.~\ref{GSE_comparison}. For the computation of the ground state, we choose an arbitrary wavefunction by choosing the parameters $\theta_i$ of the `ansatz wavefunction' from a random distribution of floating point numbers $ -2\pi \leq \theta_i < 2\pi$. Different sets of parameters are randomly chosen for the wavefunction ansatz and error-bars are computed from the standard deviation of the different results obtained.

\section{\label{packet}Wave Packet preparation and measurement}

\begin{figure*}[!ht]

\begin{minipage}{\columnwidth}
 \centering
  \includegraphics[width=\textwidth]{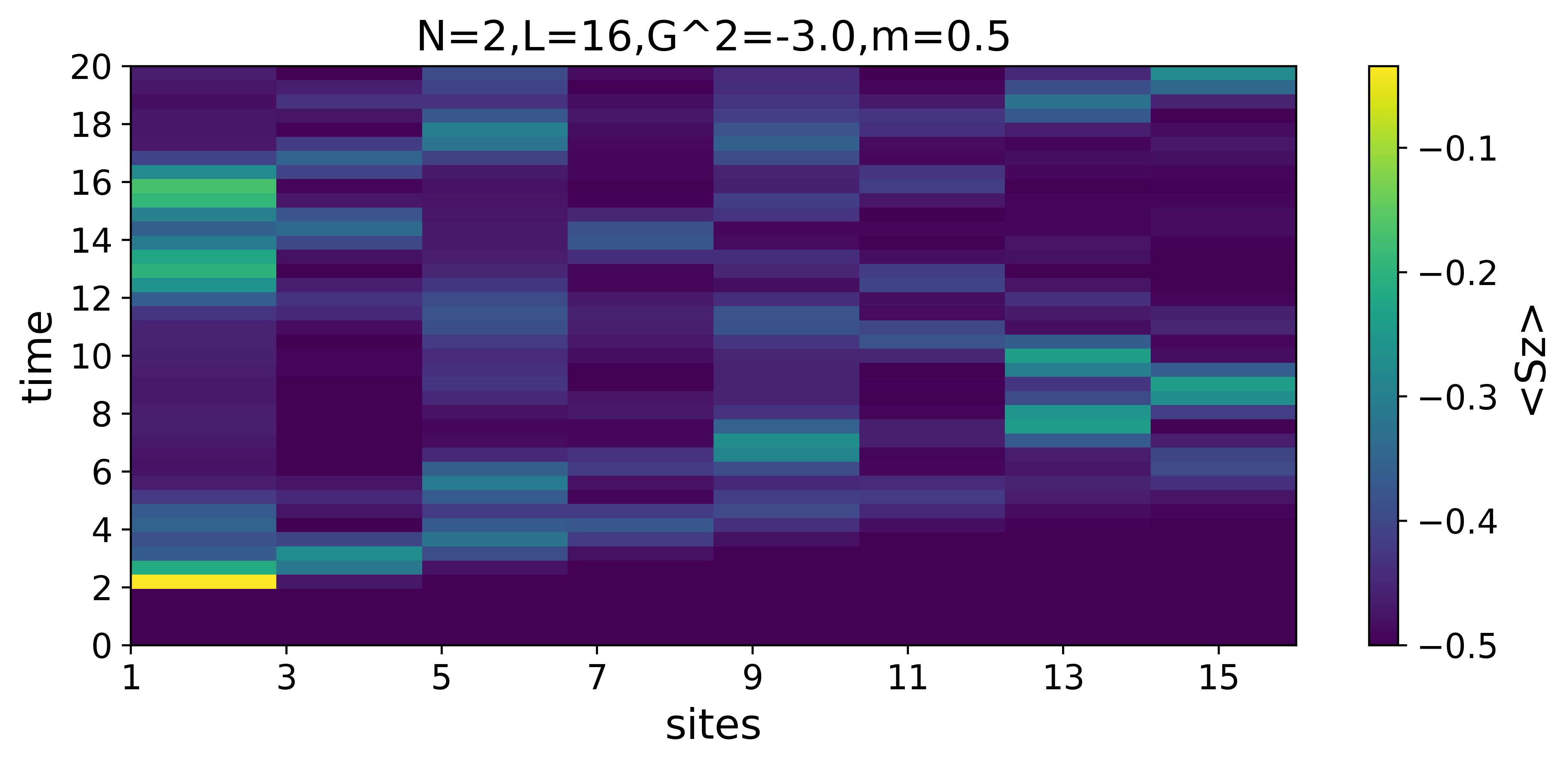}%
  \caption{Time evolution for a Right Moving Wave Packet.}
  \label{right_mover}

\end{minipage}
\begin{minipage}{\columnwidth}
 \centering
  \includegraphics[width=\textwidth]{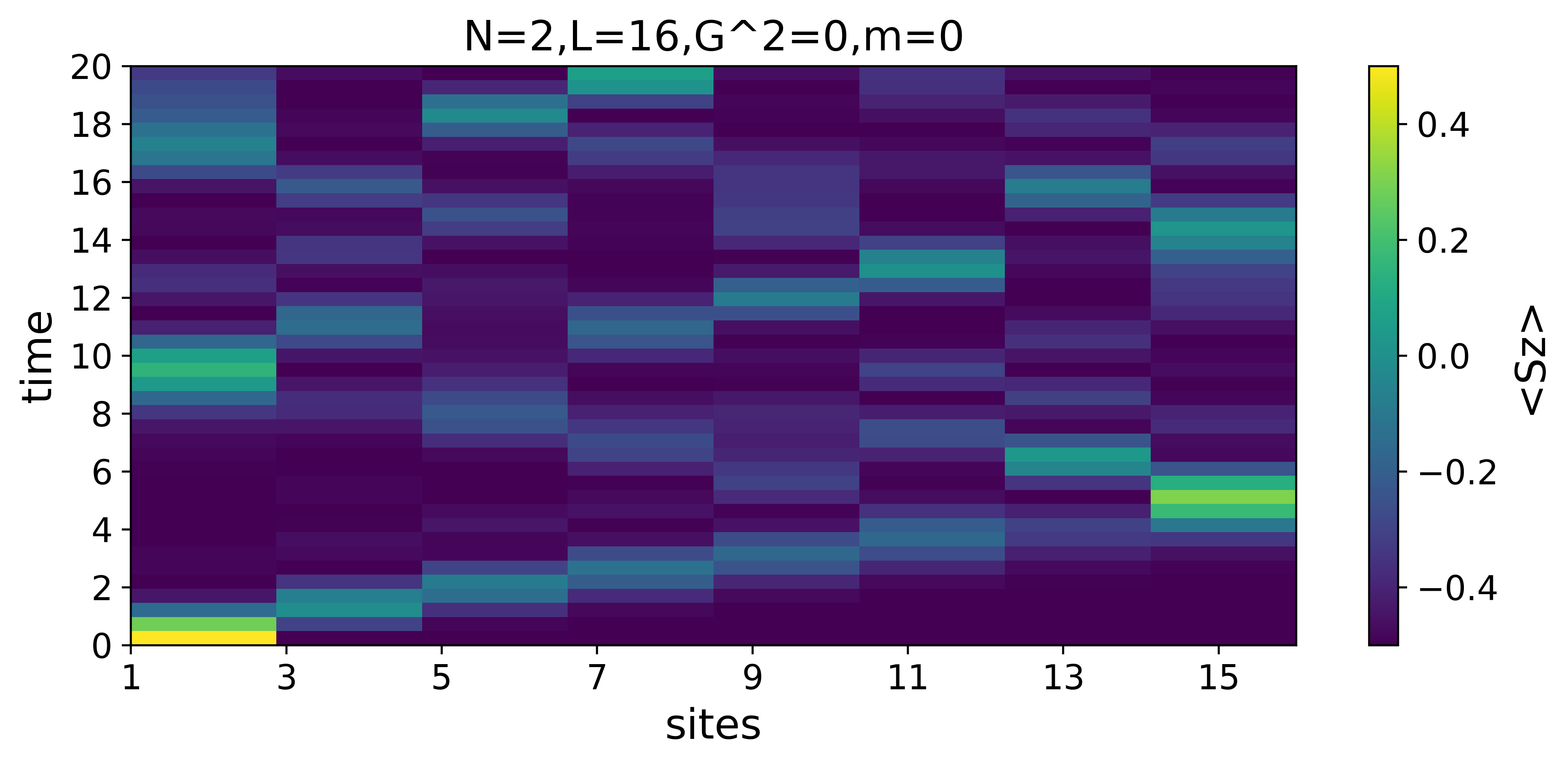}%
  \caption{Time evolution for $\ket{100\dots0}$.}
  \label{100}
\end{minipage}
\end{figure*}

\begin{figure*}[!ht]
\begin{minipage}{\columnwidth}
 \flushright
  \includegraphics[width=\textwidth]{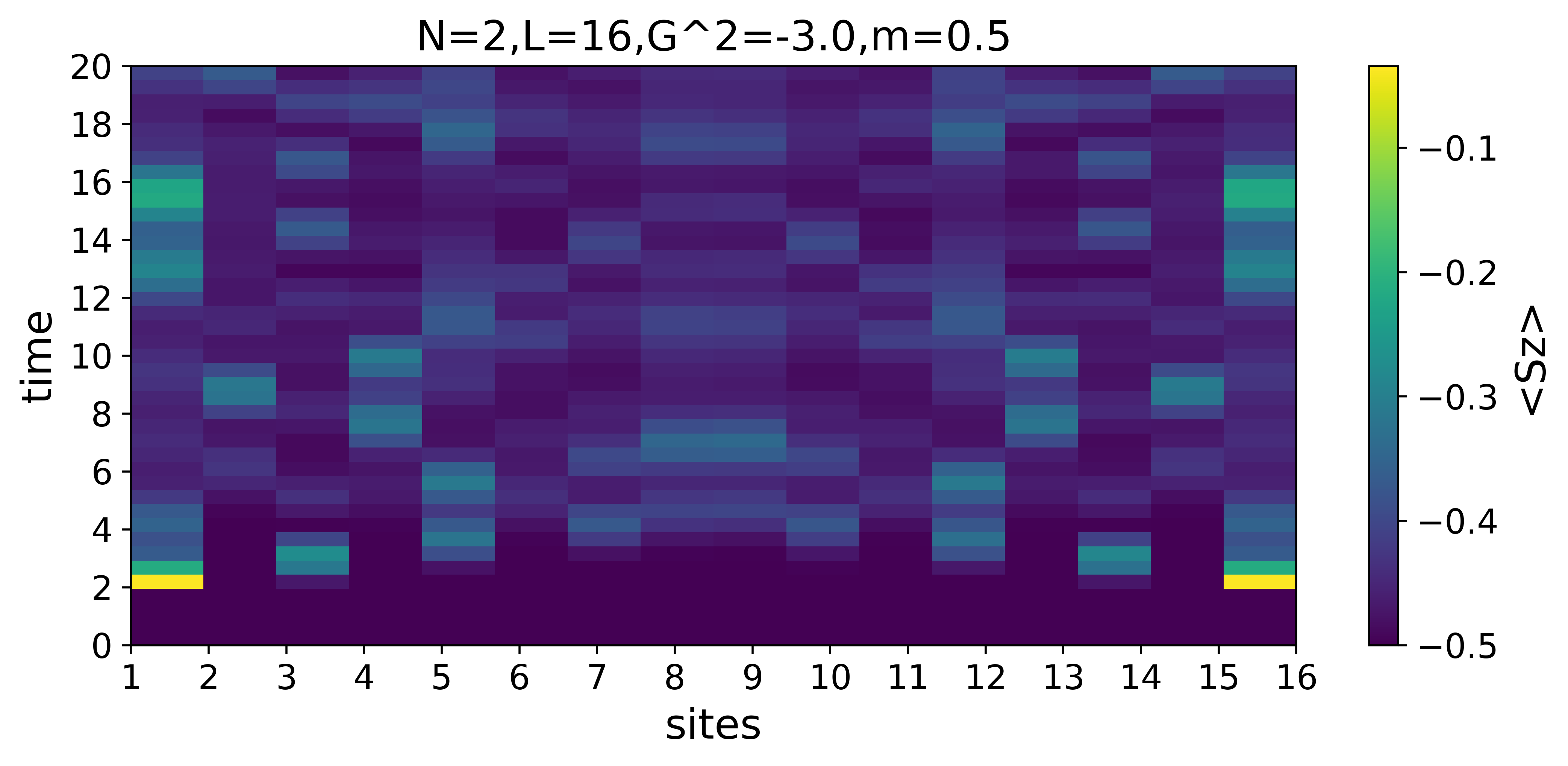}
  \caption{Time evolution for mixed Scattering State.}
  \label{scattered}
  \end{minipage}
 \begin{minipage}{\columnwidth} 
  \flushleft
  \includegraphics[width=\textwidth]{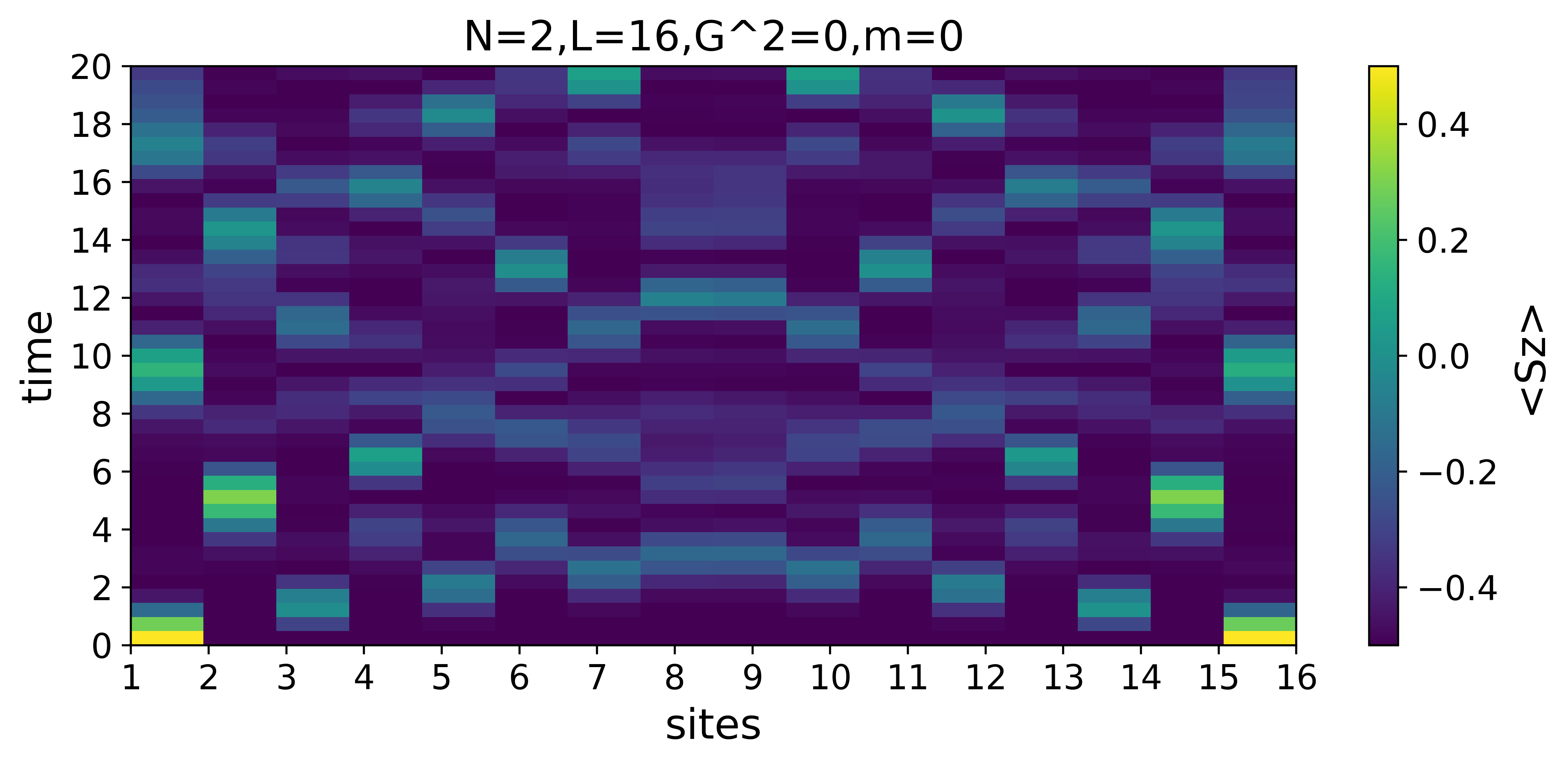}
  \caption{Time evolution for $\ket{100\dots1}$.}
  \label{101}
  \end{minipage}
\end{figure*}

In the previous three sections, we showed results of the Trotter evolution and the ground state preparation procedure for the Gross-Neveu (GN) model.
In this section, using the DMRG algorithm we demonstrate  that the Jordan, Lee and Preskill (JLP) prescription for real-time scattering can be implemented for the two flavor GN model. For the DMRG computation, we used a different labeling of the qubits with the unit
cell of the lattice containing $N$ sites corresponding to the $N$ different flavors of the model. Thus a lattice with $L$ unit cells (spatial points) would require
a total of $NL$ physical lattice sites and associated qubits. The Hamiltonian reformulated
in this notation is written in Appendix~\ref{dmrg_formulation}, see Eq.~(\ref{dmrg_action}).

The first step in constructing scattering states  is to prepare the ground state.
We obtain the ground state of the system by running the DMRG algorithm at $G^2=-3.0$, $m=0.5$ 
and obtain the ground state expressed as a Matrix Product State (MPS). This ground state is then compared with the results obtained using the variational quantum eigensolver method described in the previous section.
The ground state for any large negative $G^2$ value consists of all $\sigma_z=-1$ spins which makes it very suitable as a  starting point for wave packet creation. 
After we get the ground state in terms of an MPS  we can feed it into the TEBD algorithm as an initial state to start our time evolution for the wave packets.

Since the ground state corresponds to down states for all spins we need to add a  new term in our Hamiltonian to excite particles in precise locations on the chain (the second step in JLP prescription). The operator $
H_e = \spl(n)$ does the trick.
Note that there is no implicit sum over the site index $x$. Applying this term will change the $\ket{0}$ to a $\ket{1}$ at that location. Then we can give it some finite momentum so that this excitation can move along the spin chain: 
\begin{equation}
 H_e = e^{ik \spl(x)}.
\end{equation}
To mimic  single particle physics  we choose to excite only one site within the unit cell which results in an excitation that moves within the lattice on only even or odd lattice
sites depending on the original excited site. Since we do not want to generate particles indefinitely we will only include this term in the time evolution  for one Trotter step  and the remaining time evolution then will be carried out using the original Hamiltonian. 

 Now we give simple examples for this procedure.  In Fig.~\ref{right_mover} we have excited the first site with $k=0.5$ by including $H_e$ in the time evolution for  $1<t<2$ with a Trotter step size $\delta t=0.5$ and then let the system evolve with the original Hamiltonian up to $t=20$.  As can be seen from the plot this results in a right-moving wave packet. Notice that we have only shown the odd sites in Figs.~\ref{right_mover} and~\ref{100}.  This is due to the fact that we have excited the $1^{\rm st}$ site which resulted in an  excitation that moves only on odds sites. 
 By omitting the even sites from the plots the propagation becomes more visible. 
 Also this effectively reduces our model to be a single flavor $L=8$ model which might be accessible to NISQ era machines. 

One can also excite both ends of the spin chain to create scattering states. This can be seen in Fig.~\ref{scattered}. Due to the way we have constructed our spin chain the ends of the spin chain correspond to different flavors of fermions which results
in excitations at all lattice sites.

Alternatively we can just pick our initial state by hand to simulate wave propagation---for example by starting with a $\ket{100\dots0}$ and then time evolving this state under the original Hamiltonian with $G^2=0,m=0$. This also results in a right moving wave packet which can be seen in Fig.~\ref{100}. Furthermore  one can start with an initial state of the form $\ket{100\dots 1}$ to obtain a scattering like state in Fig.~\ref{101}. The advantage of choosing the initial state by hand and using only the kinetic term for the time evolution is that it will allow us to easily test our results with quantum hardware in the near future.

General methods to map the position space basis into a momentum space space basis for fermions have been developed in refs. \cite{ferris2014fourier,kivlichan2020improved} and can
be used to measure various aspects of the final state. Practical 
implementations with four qubits have been used to measure phase shifts using the IBM-Q and trapped ions \cite{Gustafson:2021imb} for the case of the quantum Ising model. 
Implementations with eight qubits are under active investigation.

\section{Conclusions}

In this paper, we have described a mapping of the $N$-flavor Gross Neveu (GN) model into a suitable
qubit Hamiltonian using Jordan-Wigner transformation and have benchmarked quantum simulations of the two and four flavor model by comparing its time evolution at strong coupling
using two different quantum processing units (QPUs)---the IBM-Q Guadalupe machine, which is based on superconducting transmon qubits, and the Quantinuum H1 machine, which is a trapped-ion based quantum computer.

The results of the Trotter evolution using the Guadalupe machine and the Quantinuum H1-machine are compared with the 
classical simulations, exact diagonalization, and DMRG/TEBD calculations. From the comparison with the exact computations, we find that at the current time, the Trotter evolution of the 2-flavor GN model can be reliably measured with both QPUs up-to four or five Trotter steps. However, due to the connectivity requirement of the qubits in 4-flavor model, we find that to obtain a reliable qualitative Trotter evolution, connectivity of the physical qubits in the machines becomes important. The Quantinuum H1 machine showed superior performance due to its all-to-all connectivity giving qualitatively sound results up to four Trotter steps. This indicates that configurable connectivity of the qubits is essential for implementing different multi-flavor fermionic models on current quantum machines. 

Our study also leads to a similar conclusion to some previous work  -- namely that  comparatively large Trotter steps can be used to compensate for the current limitations of current hardware which lacks quantum error correction. 
We advise the readers of an important caveat. We only used compiler optimization routines in this study for error mitigation. In our follow-up work on this model, we plan to use advanced error mitigation techniques like zero noise extrapolation. Mitigation techniques potentially can improve the 4-flavor results with the Guadalupe machine and also might improve the scaling and number of Trotter steps accessible to both platforms.

We also presented results for the ground state preparation with  a
variational quantum eigensolver. We did not attempt to implement this with quantum processing units since we suspect that ground state preparation will be impossible without advanced mitigation techniques due to the noise associated with NISQ-era machines. Even with advanced mitigation techniques, it might not be possible to prepare the ground state without improved algorithms. The authors are exploring these avenues and the results will be presented in the future. We concluded our study with wave packet preparation and scattering with DMRG/TEBD techniques, and this will be further explored with quantum processing units in the upcoming work. 

\section*{Acknowledgements}
We acknowledge useful discussions with the members of QuLat collaboration, Erik Gustafson and Bharath Sambasivam. We acknowledge support from Microsoft’s Azure Quantum team for providing credits and access to the Quantinuum hardwares and Quantinuum simulators. 
We thank the IBM-Q hub at Brookhaven National Laboratory 
for providing access to the IBMQ quantum computers. S.C, M.A and G.T were 
supported under U.S.\ Department of Energy grants DE-SC0009998 and DE-SC0019139.

\appendix

\section{Details of the DMRG Calculation \label{dmrg_formulation}}

\begin{figure}[!htb]
 \centering
  \includegraphics[height=0.30\textwidth, width=0.40\textwidth]{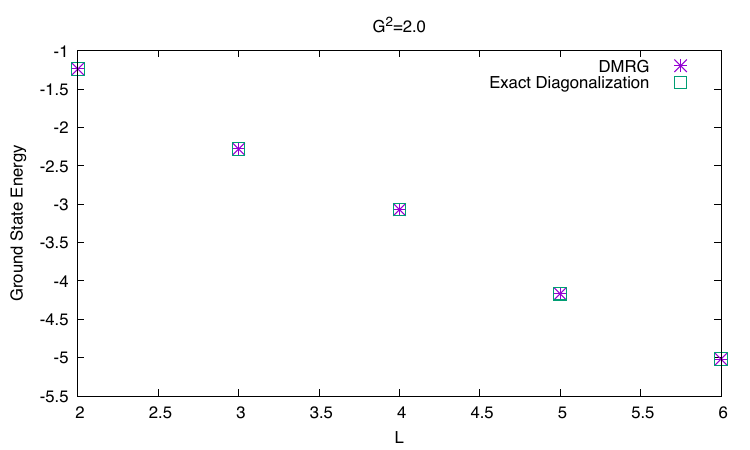}%
  \caption{Comparing the ground state energy obtained via DMRG with
  exact diagonalization for different numbers of lattice sites $L$.}
  \label{Compare_dmrg}
\end{figure}

To get a version of Eq.~\eqref{qubit_ham} suitable  for DMRG simulations \cite{PhysRevLett.69.2863,PhysRevB.48.10345,RevModPhys.77.259,itensor} 
for any number of $N$-flavors we need to make a few adjustments. Firstly we need to map
all degrees of freedom to individual lattice sites on the one dimensional
lattice.  This means the unit cell of the lattice is now of length $N$.
After this transformation the lattice action can be written as follows, 
\begin{align} \label{dmrg_action}
    H^{(N)} &= i \sum_{x=1}^{L-N} -\spl(x) \sm(x+N) +\sm(x)\spl(x+N) \nonumber \\
    &+\frac{1}{2}G^2\sum_{x=1,1+N,..}^{L-N}\sum_{a=0}^{N-1}\sum_{\substack{b=0 \\ b > a }}^{N-1}\spl(x+a)\sm(x+a) \nonumber \\
    &\times \spl(x+b)\sm(x+b).
\end{align} 
A mass term 
can be introduced using a more general staggered phase $\eta(x)$ 
that changes it's sign between the unit cells rather than between each site:
\begin{equation}
    M_s = m\sum_{x=1}^{L}\eta(x) \spl(x)\sm(x).
\end{equation}

To verify our DMRG code is correct
we have calculated the ground state energy using exact diagonalization of the 
original Hamiltonian and compared it with the energies obtained from DMRG. As can be seen
in Fig.~\ref{Compare_dmrg} there is very good agreement between the two calculations.

\section{\label{gates}Circuit Blocks}
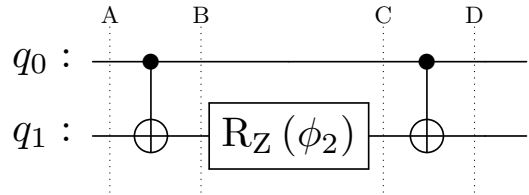
\begin{figure}[!htb]
    \centering
      \begin{tikzpicture}
        \node (mynode) [anchor=south west, inner sep=0pt] {\includegraphics[width=0.45\textwidth]{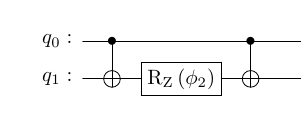}};
        \begin{scope}[x={(mynode.south east)},y={(mynode.north west)}]
          \foreach \i/\j in {{(0.3,0.8)/(0.3,0.1)},{(0.45,0.8)/(0.45,0.1)},{(0.75,0.8)/(0.75,0.1)},{(0.9,0.8)/(0.9,0.1)}}
            \draw [black, dotted] \i -- \j;
            \node[] at (0.3,0.85) {A};
            \node[] at (0.45,0.85) {B};
            \node[] at (0.75,0.85) {C};
            \node[] at (0.9,0.85) {D};
        \end{scope}
        
      \end{tikzpicture}
  \caption[short]{Circuit block for the implementation of $Q_3(\phi_2)=\exp(-i \, (\phi_2/2) \sigma_3 \otimes \sigma_3)$.}
      \label{fig_sz_sz}
  \end{figure}

For the computation of the Trotter evolution, each term in the Hamiltonian is exponentiated. This amounts to creating circuit blocks of the exponential of the tensor product of the $\sigma$ operators
$\exp{(\alpha \prod_n^\otimes \sigma_n)}$.
As an example we describe first
the creation of the operator $\exp (-i\, (\phi_{2}/2) \sigma_3 \otimes \sigma_3)$ in terms
of elementary unitary gates~\footnote{This manuscript follows the physics textbook convention, where the qubits are ordered from left to right. Thus a 2 qubit state is represented as $\ket{q_0 q_1}$, where the 0\textsuperscript{th} qubit is represented as the most significant bit in the bit string.}.
To understand the construction let us first explain the construction of the CNOT gate
\begin{equation}
    \mathrm{CNOT}_{10} = \ket{1}\bra{1} \otimes \sigma_1 + \ket{0}\bra{0} \otimes \mathbf{I}_2.
\end{equation}
Here the subscript of the $\mathrm{CNOT}$ gate, denotes that it is applied on 
qubit 1 and qubit 0 with qubit 0 as a control bit. 
We identify the basis of the one qubit states as column vectors
\begin{equation}
\ket{0}=\begin{bmatrix}
    1       \\
    0        
\end{bmatrix},
\qquad
\ket{1}=\begin{bmatrix}
    0       \\
    1        
\end{bmatrix},
\end{equation}
and $\sigma_i$, $i=1,2,3$, denotes the usual Pauli matrices.
Thus, if the initial state at position A is $\ket{\psi}$, the state at position B is $\mathrm{CNOT}_{10} \ket{\psi} $.
With these definitions, it is easy to verify that the CNOT gate applied on 2 qubits where 0\textsuperscript{th} qubit works as a control bit satisfy these identities 
\begin{align}
\mathrm{CNOT}_{10}\ket{00}&=\ket{00}, \nonumber \\
\mathrm{CNOT}_{10}\ket{01}&=\ket{01}, \nonumber \\ 
\mathrm{CNOT}_{10}\ket{10}&=\ket{11}, \nonumber \\
\mathrm{CNOT}_{10}\ket{11}&=\ket{10}.
\end{align}
Rotation by an angle $\phi_2$ around z-axis can be represented  by the rotation operator  $\mathrm{R}_z$, described by
\begin{equation}
    \mathrm{R}_z(\phi_2)=\exp\left(-i \frac{\phi_2}{2} \sigma_3\right)=\cos \frac{\phi_2}{2} \, \mathrm{I}_2 -i \, \sin \frac{\phi_2}{2} \, \sigma_3.
\end{equation}
This implies that up to the point `C' in Fig.~\ref{fig_sz_sz} the operator that is applied on a 2 qubit initial state $\ket{\psi}$ is
\begin{align}
    (\mathrm{I}_2 \otimes \mathrm{R}_z) \mathrm{CNOT}_{10} &= 
    \cos  \frac{\phi_2}{2}  \,  \mathrm{CNOT}_{10} \nonumber\\ &+\,\sin  \frac{\phi_2}{2}  (\ket{1}\bra{1} \otimes \sigma_2 + \ket{0}\bra{0} \otimes \sigma_3).
\end{align}
At the final stage at position `D', the operator takes the form
\begin{align}
    &\mathrm{CNOT}_{10}(\mathrm{I}_2 \otimes \mathrm{R}_z) \mathrm{CNOT}_{10} \nonumber \\
    &= \cos  \frac{\phi_2}{2}  \, \mathrm{I}_4  -i\,\sin  \frac{\phi_2}{2}  \big(\ket{0}\bra{0}-\ket{1}\bra{1}  \big) \otimes \sigma_3 \nonumber \\
    &=\exp(-i \,  \frac{\phi_2}{2}  \, \sigma_3 \otimes \sigma_3).
\end{align}
Thus, after the application of the second CNOT gate, the state obtained is $\exp(-i \,  (\phi_2/2)  \, \sigma_3 \otimes \sigma_3) \ket{\psi}$.  

\begin{figure}[!ht]
\centering
\includegraphics[width=0.48\textwidth]{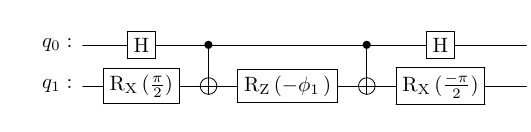}%
\caption{Circuit block for the implementation of $Q_1(-\phi_1)=\exp(i \, (\phi_1/2) \sigma_1 \otimes \sigma_2)$. \label{Q1}
}
\end{figure}

\begin{figure}[!ht]
\centering
\includegraphics[width=0.48\textwidth]{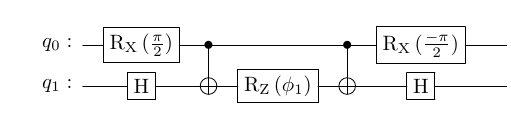}%
\caption{Circuit block for the implementation of $Q_2(\phi_1)=\exp(-i \, (\phi_1/2) \sigma_2 \otimes \sigma_1)$. \label{Q2}
}
\end{figure}

The circuit needed for $Q_1(\phi_1)$ is similar but requires Hadamard gates to
rotate the $\sigma_z$'s to $\sigma_x$ and rotation gates about the x-axis $R_x(\pi/2)$ to rotate $\sigma_z$ to  $\sigma_y$, see Fig.~\ref{Q1}. Likewise, $Q_2(-\phi_1)$ can be described by the circuit in Fig.~\ref{Q2}.

\section{\label{four}Four flavor results}

\begin{figure}[!hb]
\centering
\includegraphics[height=0.22\textwidth]{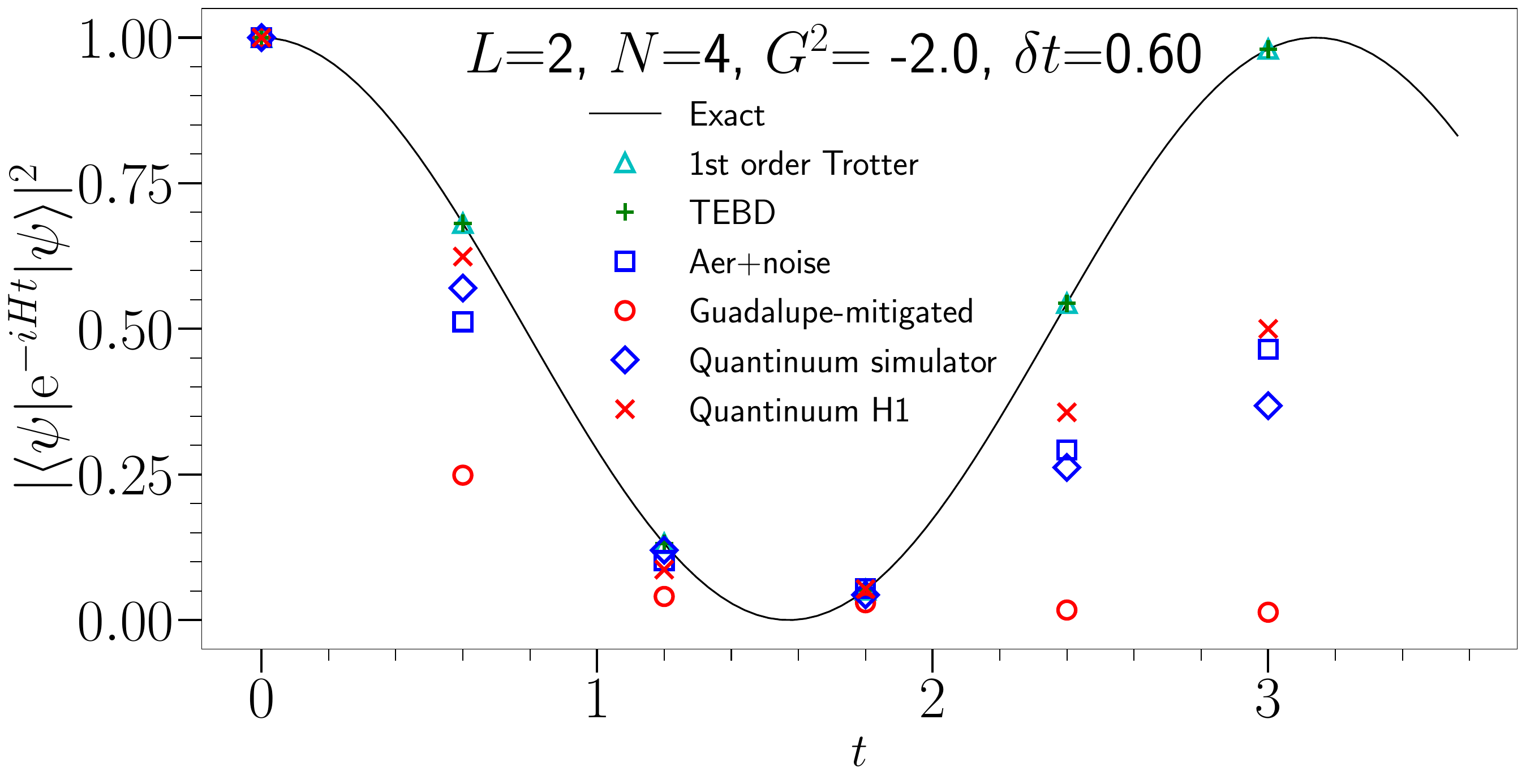}%
\caption{Trotter evolution for $N=4$ flavor model with $L=2$ lattice sites $G^2=-2.0,m=0.0$
and time step $\delta t=0.6$ from
the initial state $|\psi\rangle=|00100000\rangle$. The number of shots used for Guadalupe and Quantinuum simulations are 4000 and 300 respectively. \label{Trotter2}}
\end{figure}

In this brief section, we present the results on the Trotter evolution of 
the $N=4$ flavor Gross Neveu model. A quantum simulation demonstrates that the formulation of the $N$-flavor model is straightforward. The results obtained from the quantum circuit simulation using the Aer Simulator match with the exact diagonalization and the TEBD calculations. However, it is evident from Fig.~\ref{Trotter2} that the results obtained from the QPUs deviate from the exact results. The comparison
shows that Quantinuum's H1 QPU demonstrates superior behavior to IBMQ's Guadalupe machine. This observation is similar to that seen in the $2$-flavor model discussed in Sec. \ref{two}.  The better performance of the H1 machine can mainly be attributed to the all-to-all connectivity of its physical qubits. Due to the nature of the interaction terms, it is evident that more flavors translate to the requirement of more SWAP gates. For both cases, Trotter evolution results with QPUs deviate from the exact results more for the $N=4$ flavor case than for the $N=2$ flavor model. For example, for the Quantinuum machine, the deviation from the exact result at the fifth Trotter step is $\sim 19\%$ for two flavors, as compared to $\sim 49\%$ for the four flavor model. Whereas for the Guadalupe, the difference is $\sim 71 \%$ and $\sim 98 \%$ for the $N=2$ and $N=4$ flavor cases respectively \footnote{ The deviation is computed using the formula $\frac{|\mathrm{Exact}-QPU|}{\mathrm{Exact}}$, where $\mathrm{Exact}$ refers to the Trotter evolved result obtained using exact Trotter evolution code, whereas $\mathrm{QPU}$ refers to the results obtained from the noisy quantum processing units---guadalupe or the H1 machine.}. Furthermore, we see that the native simulator of the Quantinuum machine predicts the result quite well. In contrast, the Aer simulator with the device noise model of the Guadalupe machine does not provide an accurate description of the quantum processing unit of the Guadalupe QPU. 

\newpage

\bibliography{chiralKD}

\begin{thebibliography}{95}
\expandafter\ifx\csname natexlab\endcsname\relax\def\natexlab#1{#1}\fi
\expandafter\ifx\csname bibnamefont\endcsname\relax
  \def\bibnamefont#1{#1}\fi
\expandafter\ifx\csname bibfnamefont\endcsname\relax
  \def\bibfnamefont#1{#1}\fi
\expandafter\ifx\csname citenamefont\endcsname\relax
  \def\citenamefont#1{#1}\fi
\expandafter\ifx\csname url\endcsname\relax
  \def\url#1{\texttt{#1}}\fi
\expandafter\ifx\csname urlprefix\endcsname\relax\def\urlprefix{URL }\fi
\providecommand{\bibinfo}[2]{#2}
\providecommand{\eprint}[2][]{\url{#2}}

\bibitem[{\citenamefont{Jordan et~al.}(2014)\citenamefont{Jordan, Lee, and
  Preskill}}]{Jordan:2011ci}
\bibinfo{author}{\bibfnamefont{S.~P.} \bibnamefont{Jordan}},
  \bibinfo{author}{\bibfnamefont{K.~S.~M.} \bibnamefont{Lee}},
  \bibnamefont{and} \bibinfo{author}{\bibfnamefont{J.}~\bibnamefont{Preskill}},
  \bibinfo{journal}{Quant. Inf. Comput.} \textbf{\bibinfo{volume}{14}},
  \bibinfo{pages}{1014} (\bibinfo{year}{2014}), \eprint{1112.4833}.

\bibitem[{\citenamefont{Jordan et~al.}(2012)\citenamefont{Jordan, Lee, and
  Preskill}}]{Jordan_2012}
\bibinfo{author}{\bibfnamefont{S.~P.} \bibnamefont{Jordan}},
  \bibinfo{author}{\bibfnamefont{K.~S.~M.} \bibnamefont{Lee}},
  \bibnamefont{and} \bibinfo{author}{\bibfnamefont{J.}~\bibnamefont{Preskill}},
  \bibinfo{journal}{Science} \textbf{\bibinfo{volume}{336}},
  \bibinfo{pages}{1130} (\bibinfo{year}{2012}),
  \urlprefix\url{https://doi.org/10.1126%2Fscience.1217069}.

\bibitem[{\citenamefont{Lloyd}(1996)}]{lloyd1996universal}
\bibinfo{author}{\bibfnamefont{S.}~\bibnamefont{Lloyd}},
  \bibinfo{journal}{Science} \textbf{\bibinfo{volume}{273}},
  \bibinfo{pages}{1073} (\bibinfo{year}{1996}).

\bibitem[{\citenamefont{Mezzacapo et~al.}(2015)\citenamefont{Mezzacapo, Rico,
  Sab\'\i{}n, Egusquiza, Lamata, and Solano}}]{Mezzacapo:2015bra}
\bibinfo{author}{\bibfnamefont{A.}~\bibnamefont{Mezzacapo}},
  \bibinfo{author}{\bibfnamefont{E.}~\bibnamefont{Rico}},
  \bibinfo{author}{\bibfnamefont{C.}~\bibnamefont{Sab\'\i{}n}},
  \bibinfo{author}{\bibfnamefont{I.~L.} \bibnamefont{Egusquiza}},
  \bibinfo{author}{\bibfnamefont{L.}~\bibnamefont{Lamata}}, \bibnamefont{and}
  \bibinfo{author}{\bibfnamefont{E.}~\bibnamefont{Solano}},
  \bibinfo{journal}{Phys. Rev. Lett.} \textbf{\bibinfo{volume}{115}},
  \bibinfo{pages}{240502} (\bibinfo{year}{2015}), \eprint{1505.04720}.

\bibitem[{\citenamefont{{Cervera-Lierta}}(2018)}]{cervera2018exact}
\bibinfo{author}{\bibfnamefont{A.}~\bibnamefont{{Cervera-Lierta}}},
  \bibinfo{journal}{arXiv e-prints} \bibinfo{eid}{arXiv:1807.07112}
  (\bibinfo{year}{2018}), \eprint{1807.07112}.

\bibitem[{\citenamefont{Yeter-Aydeniz et~al.}(2019)\citenamefont{Yeter-Aydeniz,
  Dumitrescu, McCaskey, Bennink, Pooser, and Siopsis}}]{Yeter-Aydeniz:2018mix}
\bibinfo{author}{\bibfnamefont{K.}~\bibnamefont{Yeter-Aydeniz}},
  \bibinfo{author}{\bibfnamefont{E.~F.} \bibnamefont{Dumitrescu}},
  \bibinfo{author}{\bibfnamefont{A.~J.} \bibnamefont{McCaskey}},
  \bibinfo{author}{\bibfnamefont{R.~S.} \bibnamefont{Bennink}},
  \bibinfo{author}{\bibfnamefont{R.~C.} \bibnamefont{Pooser}},
  \bibnamefont{and} \bibinfo{author}{\bibfnamefont{G.}~\bibnamefont{Siopsis}},
  \bibinfo{journal}{Phys. Rev. A} \textbf{\bibinfo{volume}{99}},
  \bibinfo{pages}{032306} (\bibinfo{year}{2019}), \eprint{1811.12332}.

\bibitem[{\citenamefont{Raychowdhury and Stryker}(2020)}]{Raychowdhury:2018osk}
\bibinfo{author}{\bibfnamefont{I.}~\bibnamefont{Raychowdhury}}
  \bibnamefont{and} \bibinfo{author}{\bibfnamefont{J.~R.}
  \bibnamefont{Stryker}}, \bibinfo{journal}{Phys. Rev. Res.}
  \textbf{\bibinfo{volume}{2}}, \bibinfo{pages}{033039} (\bibinfo{year}{2020}),
  \eprint{1812.07554}.

\bibitem[{\citenamefont{Lamm and Lawrence}(2018)}]{Lamm:2018siq}
\bibinfo{author}{\bibfnamefont{H.}~\bibnamefont{Lamm}} \bibnamefont{and}
  \bibinfo{author}{\bibfnamefont{S.}~\bibnamefont{Lawrence}},
  \bibinfo{journal}{Phys. Rev. Lett.} \textbf{\bibinfo{volume}{121}},
  \bibinfo{pages}{170501} (\bibinfo{year}{2018}), \eprint{1806.06649}.

\bibitem[{\citenamefont{Macridin et~al.}(2018)\citenamefont{Macridin,
  Spentzouris, Amundson, and Harnik}}]{macridin2018digital}
\bibinfo{author}{\bibfnamefont{A.}~\bibnamefont{Macridin}},
  \bibinfo{author}{\bibfnamefont{P.}~\bibnamefont{Spentzouris}},
  \bibinfo{author}{\bibfnamefont{J.}~\bibnamefont{Amundson}}, \bibnamefont{and}
  \bibinfo{author}{\bibfnamefont{R.}~\bibnamefont{Harnik}},
  \bibinfo{journal}{Phys. Rev. A} \textbf{\bibinfo{volume}{98}},
  \bibinfo{pages}{042312} (\bibinfo{year}{2018}), \eprint{1805.09928}.

\bibitem[{\citenamefont{Klco et~al.}(2018)\citenamefont{Klco, Dumitrescu,
  McCaskey, Morris, Pooser, Sanz, Solano, Lougovski, and
  Savage}}]{Klco:2018kyo}
\bibinfo{author}{\bibfnamefont{N.}~\bibnamefont{Klco}},
  \bibinfo{author}{\bibfnamefont{E.~F.} \bibnamefont{Dumitrescu}},
  \bibinfo{author}{\bibfnamefont{A.~J.} \bibnamefont{McCaskey}},
  \bibinfo{author}{\bibfnamefont{T.~D.} \bibnamefont{Morris}},
  \bibinfo{author}{\bibfnamefont{R.~C.} \bibnamefont{Pooser}},
  \bibinfo{author}{\bibfnamefont{M.}~\bibnamefont{Sanz}},
  \bibinfo{author}{\bibfnamefont{E.}~\bibnamefont{Solano}},
  \bibinfo{author}{\bibfnamefont{P.}~\bibnamefont{Lougovski}},
  \bibnamefont{and} \bibinfo{author}{\bibfnamefont{M.~J.}
  \bibnamefont{Savage}}, \bibinfo{journal}{Phys. Rev. A}
  \textbf{\bibinfo{volume}{98}}, \bibinfo{pages}{032331}
  (\bibinfo{year}{2018}), \eprint{1803.03326}.

\bibitem[{\citenamefont{Bauer et~al.}(2021)\citenamefont{Bauer, de~Jong,
  Nachman, and Provasoli}}]{Bauer:2019qxa}
\bibinfo{author}{\bibfnamefont{C.~W.} \bibnamefont{Bauer}},
  \bibinfo{author}{\bibfnamefont{W.~A.} \bibnamefont{de~Jong}},
  \bibinfo{author}{\bibfnamefont{B.}~\bibnamefont{Nachman}}, \bibnamefont{and}
  \bibinfo{author}{\bibfnamefont{D.}~\bibnamefont{Provasoli}},
  \bibinfo{journal}{Phys. Rev. Lett.} \textbf{\bibinfo{volume}{126}},
  \bibinfo{pages}{062001} (\bibinfo{year}{2021}), \eprint{1904.03196}.

\bibitem[{\citenamefont{Lamm et~al.}(2020)\citenamefont{Lamm, Lawrence, and
  Yamauchi}}]{Lamm:2019uyc}
\bibinfo{author}{\bibfnamefont{H.}~\bibnamefont{Lamm}},
  \bibinfo{author}{\bibfnamefont{S.}~\bibnamefont{Lawrence}}, \bibnamefont{and}
  \bibinfo{author}{\bibfnamefont{Y.}~\bibnamefont{Yamauchi}}
  (\bibinfo{collaboration}{NuQS}), \bibinfo{journal}{Phys. Rev. Res.}
  \textbf{\bibinfo{volume}{2}}, \bibinfo{pages}{013272} (\bibinfo{year}{2020}),
  \eprint{1908.10439}.

\bibitem[{\citenamefont{Lamm et~al.}(2019)\citenamefont{Lamm, Lawrence, and
  Yamauchi}}]{Lamm:2019bik}
\bibinfo{author}{\bibfnamefont{H.}~\bibnamefont{Lamm}},
  \bibinfo{author}{\bibfnamefont{S.}~\bibnamefont{Lawrence}}, \bibnamefont{and}
  \bibinfo{author}{\bibfnamefont{Y.}~\bibnamefont{Yamauchi}}
  (\bibinfo{collaboration}{NuQS}), \bibinfo{journal}{Phys. Rev. D}
  \textbf{\bibinfo{volume}{100}}, \bibinfo{pages}{034518}
  (\bibinfo{year}{2019}), \eprint{1903.08807}.

\bibitem[{\citenamefont{Gustafson et~al.}(2019)\citenamefont{Gustafson,
  Meurice, and Unmuth-Yockey}}]{Gustafson:2019mpk}
\bibinfo{author}{\bibfnamefont{E.}~\bibnamefont{Gustafson}},
  \bibinfo{author}{\bibfnamefont{Y.}~\bibnamefont{Meurice}}, \bibnamefont{and}
  \bibinfo{author}{\bibfnamefont{J.}~\bibnamefont{Unmuth-Yockey}},
  \bibinfo{journal}{Phys. Rev. D} \textbf{\bibinfo{volume}{99}},
  \bibinfo{pages}{094503} (\bibinfo{year}{2019}), \eprint{1901.05944}.

\bibitem[{\citenamefont{Gustafson
  et~al.}(2021{\natexlab{a}})\citenamefont{Gustafson, Dreher, Hang, and
  Meurice}}]{gustafson2021indexed}
\bibinfo{author}{\bibfnamefont{E.}~\bibnamefont{Gustafson}},
  \bibinfo{author}{\bibfnamefont{P.}~\bibnamefont{Dreher}},
  \bibinfo{author}{\bibfnamefont{Z.}~\bibnamefont{Hang}}, \bibnamefont{and}
  \bibinfo{author}{\bibfnamefont{Y.}~\bibnamefont{Meurice}},
  \bibinfo{journal}{Quantum Science and Technology}
  \textbf{\bibinfo{volume}{6}}, \bibinfo{pages}{045020}
  (\bibinfo{year}{2021}{\natexlab{a}}).

\bibitem[{\citenamefont{Gustafson
  et~al.}(2021{\natexlab{b}})\citenamefont{Gustafson, Zhu, Dreher, Linke, and
  Meurice}}]{Gustafson:2021imb}
\bibinfo{author}{\bibfnamefont{E.}~\bibnamefont{Gustafson}},
  \bibinfo{author}{\bibfnamefont{Y.}~\bibnamefont{Zhu}},
  \bibinfo{author}{\bibfnamefont{P.}~\bibnamefont{Dreher}},
  \bibinfo{author}{\bibfnamefont{N.~M.} \bibnamefont{Linke}}, \bibnamefont{and}
  \bibinfo{author}{\bibfnamefont{Y.}~\bibnamefont{Meurice}},
  \bibinfo{journal}{Phys. Rev. D} \textbf{\bibinfo{volume}{104}},
  \bibinfo{pages}{054507} (\bibinfo{year}{2021}{\natexlab{b}}),
  \eprint{2103.06848}.

\bibitem[{\citenamefont{Kharzeev and Kikuchi}(2020)}]{Kharzeev:2020kgc}
\bibinfo{author}{\bibfnamefont{D.~E.} \bibnamefont{Kharzeev}} \bibnamefont{and}
  \bibinfo{author}{\bibfnamefont{Y.}~\bibnamefont{Kikuchi}},
  \bibinfo{journal}{Phys. Rev. Res.} \textbf{\bibinfo{volume}{2}},
  \bibinfo{pages}{023342} (\bibinfo{year}{2020}), \eprint{2001.00698}.

\bibitem[{\citenamefont{Honda et~al.}(2022)\citenamefont{Honda, Itou, Kikuchi,
  Nagano, and Okuda}}]{Honda:2021aum}
\bibinfo{author}{\bibfnamefont{M.}~\bibnamefont{Honda}},
  \bibinfo{author}{\bibfnamefont{E.}~\bibnamefont{Itou}},
  \bibinfo{author}{\bibfnamefont{Y.}~\bibnamefont{Kikuchi}},
  \bibinfo{author}{\bibfnamefont{L.}~\bibnamefont{Nagano}}, \bibnamefont{and}
  \bibinfo{author}{\bibfnamefont{T.}~\bibnamefont{Okuda}},
  \bibinfo{journal}{Phys. Rev. D} \textbf{\bibinfo{volume}{105}},
  \bibinfo{pages}{014504} (\bibinfo{year}{2022}), \eprint{2105.03276}.

\bibitem[{\citenamefont{Bhattacharya et~al.}(2021)\citenamefont{Bhattacharya,
  Buser, Chandrasekharan, Gupta, and Singh}}]{Bhattacharya:2020gpm}
\bibinfo{author}{\bibfnamefont{T.}~\bibnamefont{Bhattacharya}},
  \bibinfo{author}{\bibfnamefont{A.~J.} \bibnamefont{Buser}},
  \bibinfo{author}{\bibfnamefont{S.}~\bibnamefont{Chandrasekharan}},
  \bibinfo{author}{\bibfnamefont{R.}~\bibnamefont{Gupta}}, \bibnamefont{and}
  \bibinfo{author}{\bibfnamefont{H.}~\bibnamefont{Singh}},
  \bibinfo{journal}{Phys. Rev. Lett.} \textbf{\bibinfo{volume}{126}},
  \bibinfo{pages}{172001} (\bibinfo{year}{2021}), \eprint{2012.02153}.

\bibitem[{\citenamefont{Ji et~al.}(2022)\citenamefont{Ji, Lamm, and
  Zhu}}]{Ji:2022qvr}
\bibinfo{author}{\bibfnamefont{Y.}~\bibnamefont{Ji}},
  \bibinfo{author}{\bibfnamefont{H.}~\bibnamefont{Lamm}}, \bibnamefont{and}
  \bibinfo{author}{\bibfnamefont{S.}~\bibnamefont{Zhu}} (\bibinfo{year}{2022}),
  \eprint{2203.02330}.

\bibitem[{\citenamefont{Bloch et~al.}(2012)\citenamefont{Bloch, Dalibard, and
  Nascimbene}}]{bloch2012quantum}
\bibinfo{author}{\bibfnamefont{I.}~\bibnamefont{Bloch}},
  \bibinfo{author}{\bibfnamefont{J.}~\bibnamefont{Dalibard}}, \bibnamefont{and}
  \bibinfo{author}{\bibfnamefont{S.}~\bibnamefont{Nascimbene}},
  \bibinfo{journal}{Nature Physics} \textbf{\bibinfo{volume}{8}},
  \bibinfo{pages}{267} (\bibinfo{year}{2012}).

\bibitem[{\citenamefont{Lewenstein et~al.}(2007)\citenamefont{Lewenstein,
  Sanpera, Ahufinger, Damski, Sen, and Sen}}]{lewenstein2007ultracold}
\bibinfo{author}{\bibfnamefont{M.}~\bibnamefont{Lewenstein}},
  \bibinfo{author}{\bibfnamefont{A.}~\bibnamefont{Sanpera}},
  \bibinfo{author}{\bibfnamefont{V.}~\bibnamefont{Ahufinger}},
  \bibinfo{author}{\bibfnamefont{B.}~\bibnamefont{Damski}},
  \bibinfo{author}{\bibfnamefont{A.}~\bibnamefont{Sen}}, \bibnamefont{and}
  \bibinfo{author}{\bibfnamefont{U.}~\bibnamefont{Sen}},
  \bibinfo{journal}{Advances in Physics} \textbf{\bibinfo{volume}{56}},
  \bibinfo{pages}{243} (\bibinfo{year}{2007}).

\bibitem[{\citenamefont{Cirac et~al.}(2010)\citenamefont{Cirac, Maraner, and
  Pachos}}]{cirac2010cold}
\bibinfo{author}{\bibfnamefont{J.~I.} \bibnamefont{Cirac}},
  \bibinfo{author}{\bibfnamefont{P.}~\bibnamefont{Maraner}}, \bibnamefont{and}
  \bibinfo{author}{\bibfnamefont{J.~K.} \bibnamefont{Pachos}},
  \bibinfo{journal}{Physical review letters} \textbf{\bibinfo{volume}{105}},
  \bibinfo{pages}{190403} (\bibinfo{year}{2010}).

\bibitem[{\citenamefont{Kapit and Mueller}(2011)}]{Kapit:2010qu}
\bibinfo{author}{\bibfnamefont{E.}~\bibnamefont{Kapit}} \bibnamefont{and}
  \bibinfo{author}{\bibfnamefont{E.~J.} \bibnamefont{Mueller}},
  \bibinfo{journal}{Phys. Rev. A} \textbf{\bibinfo{volume}{83}},
  \bibinfo{pages}{033625} (\bibinfo{year}{2011}), \eprint{1011.4021}.

\bibitem[{\citenamefont{Kuno et~al.}(2017)\citenamefont{Kuno, Sakane,
  Kasamatsu, Ichinose, and Matsui}}]{Kuno:2016ipi}
\bibinfo{author}{\bibfnamefont{Y.}~\bibnamefont{Kuno}},
  \bibinfo{author}{\bibfnamefont{S.}~\bibnamefont{Sakane}},
  \bibinfo{author}{\bibfnamefont{K.}~\bibnamefont{Kasamatsu}},
  \bibinfo{author}{\bibfnamefont{I.}~\bibnamefont{Ichinose}}, \bibnamefont{and}
  \bibinfo{author}{\bibfnamefont{T.}~\bibnamefont{Matsui}},
  \bibinfo{journal}{Phys. Rev. D} \textbf{\bibinfo{volume}{95}},
  \bibinfo{pages}{094507} (\bibinfo{year}{2017}), \eprint{1605.00333}.

\bibitem[{\citenamefont{Martinez et~al.}(2016)}]{Martinez:2016yna}
\bibinfo{author}{\bibfnamefont{E.~A.} \bibnamefont{Martinez}}
  \bibnamefont{et~al.}, \bibinfo{journal}{Nature}
  \textbf{\bibinfo{volume}{534}}, \bibinfo{pages}{516} (\bibinfo{year}{2016}),
  \eprint{1605.04570}.

\bibitem[{\citenamefont{Danshita et~al.}(2017)\citenamefont{Danshita, Hanada,
  and Tezuka}}]{Danshita:2016xbo}
\bibinfo{author}{\bibfnamefont{I.}~\bibnamefont{Danshita}},
  \bibinfo{author}{\bibfnamefont{M.}~\bibnamefont{Hanada}}, \bibnamefont{and}
  \bibinfo{author}{\bibfnamefont{M.}~\bibnamefont{Tezuka}},
  \bibinfo{journal}{PTEP} \textbf{\bibinfo{volume}{2017}},
  \bibinfo{pages}{083I01} (\bibinfo{year}{2017}), \eprint{1606.02454}.

\bibitem[{\citenamefont{Zhang et~al.}(2018)\citenamefont{Zhang, Unmuth-Yockey,
  Zeiher, Bazavov, Tsai, and Meurice}}]{Zhang:2018ufj}
\bibinfo{author}{\bibfnamefont{J.}~\bibnamefont{Zhang}},
  \bibinfo{author}{\bibfnamefont{J.}~\bibnamefont{Unmuth-Yockey}},
  \bibinfo{author}{\bibfnamefont{J.}~\bibnamefont{Zeiher}},
  \bibinfo{author}{\bibfnamefont{A.}~\bibnamefont{Bazavov}},
  \bibinfo{author}{\bibfnamefont{S.~W.} \bibnamefont{Tsai}}, \bibnamefont{and}
  \bibinfo{author}{\bibfnamefont{Y.}~\bibnamefont{Meurice}},
  \bibinfo{journal}{Phys. Rev. Lett.} \textbf{\bibinfo{volume}{121}},
  \bibinfo{pages}{223201} (\bibinfo{year}{2018}), \eprint{1803.11166}.

\bibitem[{\citenamefont{Davoudi et~al.}(2021)\citenamefont{Davoudi, Linke, and
  Pagano}}]{Davoudi:2021ney}
\bibinfo{author}{\bibfnamefont{Z.}~\bibnamefont{Davoudi}},
  \bibinfo{author}{\bibfnamefont{N.~M.} \bibnamefont{Linke}}, \bibnamefont{and}
  \bibinfo{author}{\bibfnamefont{G.}~\bibnamefont{Pagano}},
  \bibinfo{journal}{Phys. Rev. Res.} \textbf{\bibinfo{volume}{3}},
  \bibinfo{pages}{043072} (\bibinfo{year}{2021}), \eprint{2104.09346}.

\bibitem[{\citenamefont{Davoudi et~al.}(2020)\citenamefont{Davoudi, Hafezi,
  Monroe, Pagano, Seif, and Shaw}}]{Davoudi:2019bhy}
\bibinfo{author}{\bibfnamefont{Z.}~\bibnamefont{Davoudi}},
  \bibinfo{author}{\bibfnamefont{M.}~\bibnamefont{Hafezi}},
  \bibinfo{author}{\bibfnamefont{C.}~\bibnamefont{Monroe}},
  \bibinfo{author}{\bibfnamefont{G.}~\bibnamefont{Pagano}},
  \bibinfo{author}{\bibfnamefont{A.}~\bibnamefont{Seif}}, \bibnamefont{and}
  \bibinfo{author}{\bibfnamefont{A.}~\bibnamefont{Shaw}},
  \bibinfo{journal}{Phys. Rev. Res.} \textbf{\bibinfo{volume}{2}},
  \bibinfo{pages}{023015} (\bibinfo{year}{2020}), \eprint{1908.03210}.

\bibitem[{\citenamefont{Monroe et~al.}(2021)}]{Monroe:2019asq}
\bibinfo{author}{\bibfnamefont{C.}~\bibnamefont{Monroe}} \bibnamefont{et~al.},
  \bibinfo{journal}{Rev. Mod. Phys.} \textbf{\bibinfo{volume}{93}},
  \bibinfo{pages}{025001} (\bibinfo{year}{2021}), \eprint{1912.07845}.

\bibitem[{\citenamefont{Gonz\'alez-Cuadra
  et~al.}(2017)\citenamefont{Gonz\'alez-Cuadra, Zohar, and
  Cirac}}]{Gonzalez-Cuadra:2017lvz}
\bibinfo{author}{\bibfnamefont{D.}~\bibnamefont{Gonz\'alez-Cuadra}},
  \bibinfo{author}{\bibfnamefont{E.}~\bibnamefont{Zohar}}, \bibnamefont{and}
  \bibinfo{author}{\bibfnamefont{J.~I.} \bibnamefont{Cirac}},
  \bibinfo{journal}{New J. Phys.} \textbf{\bibinfo{volume}{19}},
  \bibinfo{pages}{063038} (\bibinfo{year}{2017}), \eprint{1702.05492}.

\bibitem[{\citenamefont{Nguyen et~al.}(2022)\citenamefont{Nguyen, Tran, Zhu,
  Green, Alderete, Davoudi, and Linke}}]{Nguyen:2021hyk}
\bibinfo{author}{\bibfnamefont{N.~H.} \bibnamefont{Nguyen}},
  \bibinfo{author}{\bibfnamefont{M.~C.} \bibnamefont{Tran}},
  \bibinfo{author}{\bibfnamefont{Y.}~\bibnamefont{Zhu}},
  \bibinfo{author}{\bibfnamefont{A.~M.} \bibnamefont{Green}},
  \bibinfo{author}{\bibfnamefont{C.~H.} \bibnamefont{Alderete}},
  \bibinfo{author}{\bibfnamefont{Z.}~\bibnamefont{Davoudi}}, \bibnamefont{and}
  \bibinfo{author}{\bibfnamefont{N.~M.} \bibnamefont{Linke}},
  \bibinfo{journal}{PRX Quantum} \textbf{\bibinfo{volume}{3}},
  \bibinfo{pages}{020324} (\bibinfo{year}{2022}), \eprint{2112.14262}.

\bibitem[{\citenamefont{Aidelsburger et~al.}(2021)}]{Aidelsburger:2021mia}
\bibinfo{author}{\bibfnamefont{M.}~\bibnamefont{Aidelsburger}}
  \bibnamefont{et~al.}, \bibinfo{journal}{Phil. Trans. Roy. Soc. Lond. A}
  \textbf{\bibinfo{volume}{380}}, \bibinfo{pages}{20210064}
  (\bibinfo{year}{2021}), \eprint{2106.03063}.

\bibitem[{\citenamefont{Schweizer et~al.}(2019)\citenamefont{Schweizer, Grusdt,
  Berngruber, Barbiero, Demler, Goldman, Bloch, and
  Aidelsburger}}]{Schweizer:2019lwx}
\bibinfo{author}{\bibfnamefont{C.}~\bibnamefont{Schweizer}},
  \bibinfo{author}{\bibfnamefont{F.}~\bibnamefont{Grusdt}},
  \bibinfo{author}{\bibfnamefont{M.}~\bibnamefont{Berngruber}},
  \bibinfo{author}{\bibfnamefont{L.}~\bibnamefont{Barbiero}},
  \bibinfo{author}{\bibfnamefont{E.}~\bibnamefont{Demler}},
  \bibinfo{author}{\bibfnamefont{N.}~\bibnamefont{Goldman}},
  \bibinfo{author}{\bibfnamefont{I.}~\bibnamefont{Bloch}}, \bibnamefont{and}
  \bibinfo{author}{\bibfnamefont{M.}~\bibnamefont{Aidelsburger}},
  \bibinfo{journal}{Nature Phys.} \textbf{\bibinfo{volume}{15}},
  \bibinfo{pages}{1168} (\bibinfo{year}{2019}), \eprint{1901.07103}.

\bibitem[{\citenamefont{Bauer et~al.}(2022)}]{Bauer:2022hpo}
\bibinfo{author}{\bibfnamefont{C.~W.} \bibnamefont{Bauer}} \bibnamefont{et~al.}
  (\bibinfo{year}{2022}), \eprint{2204.03381}.

\bibitem[{\citenamefont{Ba\~nuls et~al.}(2020)}]{Banuls:2019bmf}
\bibinfo{author}{\bibfnamefont{M.~C.} \bibnamefont{Ba\~nuls}}
  \bibnamefont{et~al.}, \bibinfo{journal}{Eur. Phys. J. D}
  \textbf{\bibinfo{volume}{74}}, \bibinfo{pages}{165} (\bibinfo{year}{2020}),
  \eprint{1911.00003}.

\bibitem[{\citenamefont{Kasper et~al.}(2020)\citenamefont{Kasper, Juzeliunas,
  Lewenstein, Jendrzejewski, and Zohar}}]{Kasper:2020akk}
\bibinfo{author}{\bibfnamefont{V.}~\bibnamefont{Kasper}},
  \bibinfo{author}{\bibfnamefont{G.}~\bibnamefont{Juzeliunas}},
  \bibinfo{author}{\bibfnamefont{M.}~\bibnamefont{Lewenstein}},
  \bibinfo{author}{\bibfnamefont{F.}~\bibnamefont{Jendrzejewski}},
  \bibnamefont{and} \bibinfo{author}{\bibfnamefont{E.}~\bibnamefont{Zohar}},
  \bibinfo{journal}{New J. Phys.} \textbf{\bibinfo{volume}{22}},
  \bibinfo{pages}{103027} (\bibinfo{year}{2020}), \eprint{2006.01258}.

\bibitem[{\citenamefont{Dalmonte and Montangero}(2016)}]{Dalmonte:2016alw}
\bibinfo{author}{\bibfnamefont{M.}~\bibnamefont{Dalmonte}} \bibnamefont{and}
  \bibinfo{author}{\bibfnamefont{S.}~\bibnamefont{Montangero}},
  \bibinfo{journal}{Contemp. Phys.} \textbf{\bibinfo{volume}{57}},
  \bibinfo{pages}{388} (\bibinfo{year}{2016}), \eprint{1602.03776}.

\bibitem[{\citenamefont{Meurice et~al.}(2022)\citenamefont{Meurice, Sakai, and
  Unmuth-Yockey}}]{meurice2022tensor}
\bibinfo{author}{\bibfnamefont{Y.}~\bibnamefont{Meurice}},
  \bibinfo{author}{\bibfnamefont{R.}~\bibnamefont{Sakai}}, \bibnamefont{and}
  \bibinfo{author}{\bibfnamefont{J.}~\bibnamefont{Unmuth-Yockey}},
  \bibinfo{journal}{Rev. Mod. Phys.} \textbf{\bibinfo{volume}{94}},
  \bibinfo{pages}{025005} (\bibinfo{year}{2022}), \eprint{2010.06539},
  \urlprefix\url{https://link.aps.org/doi/10.1103/RevModPhys.94.025005}.

\bibitem[{\citenamefont{Kogut}(1979)}]{RevModPhys.51.659}
\bibinfo{author}{\bibfnamefont{J.~B.} \bibnamefont{Kogut}},
  \bibinfo{journal}{Rev. Mod. Phys.} \textbf{\bibinfo{volume}{51}},
  \bibinfo{pages}{659} (\bibinfo{year}{1979}),
  \urlprefix\url{https://link.aps.org/doi/10.1103/RevModPhys.51.659}.

\bibitem[{\citenamefont{Hauke et~al.}(2013)\citenamefont{Hauke, Marcos,
  Dalmonte, and Zoller}}]{Hauke:2013jga}
\bibinfo{author}{\bibfnamefont{P.}~\bibnamefont{Hauke}},
  \bibinfo{author}{\bibfnamefont{D.}~\bibnamefont{Marcos}},
  \bibinfo{author}{\bibfnamefont{M.}~\bibnamefont{Dalmonte}}, \bibnamefont{and}
  \bibinfo{author}{\bibfnamefont{P.}~\bibnamefont{Zoller}},
  \bibinfo{journal}{Phys. Rev. X} \textbf{\bibinfo{volume}{3}},
  \bibinfo{pages}{041018} (\bibinfo{year}{2013}), \eprint{1306.2162}.

\bibitem[{\citenamefont{K\"uhn et~al.}(2014)\citenamefont{K\"uhn, Cirac, and
  Ba\~nuls}}]{Kuhn:2014rha}
\bibinfo{author}{\bibfnamefont{S.}~\bibnamefont{K\"uhn}},
  \bibinfo{author}{\bibfnamefont{J.~I.} \bibnamefont{Cirac}}, \bibnamefont{and}
  \bibinfo{author}{\bibfnamefont{M.-C.} \bibnamefont{Ba\~nuls}},
  \bibinfo{journal}{Phys. Rev. A} \textbf{\bibinfo{volume}{90}},
  \bibinfo{pages}{042305} (\bibinfo{year}{2014}), \eprint{1407.4995}.

\bibitem[{\citenamefont{Thompson and Siopsis}(2022)}]{Thompson:2021eze}
\bibinfo{author}{\bibfnamefont{S.}~\bibnamefont{Thompson}} \bibnamefont{and}
  \bibinfo{author}{\bibfnamefont{G.}~\bibnamefont{Siopsis}},
  \bibinfo{journal}{Quantum Sci. Technol.} \textbf{\bibinfo{volume}{7}},
  \bibinfo{pages}{035001} (\bibinfo{year}{2022}), \eprint{2110.13046}.

\bibitem[{\citenamefont{Shaw et~al.}(2020)\citenamefont{Shaw, Lougovski,
  Stryker, and Wiebe}}]{Shaw:2020udc}
\bibinfo{author}{\bibfnamefont{A.~F.} \bibnamefont{Shaw}},
  \bibinfo{author}{\bibfnamefont{P.}~\bibnamefont{Lougovski}},
  \bibinfo{author}{\bibfnamefont{J.~R.} \bibnamefont{Stryker}},
  \bibnamefont{and} \bibinfo{author}{\bibfnamefont{N.}~\bibnamefont{Wiebe}},
  \bibinfo{journal}{Quantum} \textbf{\bibinfo{volume}{4}}, \bibinfo{pages}{306}
  (\bibinfo{year}{2020}), \eprint{2002.11146}.

\bibitem[{\citenamefont{Gross and Neveu}(1974)}]{Gross:1974jv}
\bibinfo{author}{\bibfnamefont{D.~J.} \bibnamefont{Gross}} \bibnamefont{and}
  \bibinfo{author}{\bibfnamefont{A.}~\bibnamefont{Neveu}},
  \bibinfo{journal}{Phys. Rev. D} \textbf{\bibinfo{volume}{10}},
  \bibinfo{pages}{3235} (\bibinfo{year}{1974}).

\bibitem[{\citenamefont{Hamed~Moosavian and
  Jordan}(2018)}]{HamedMoosavian:2017koz}
\bibinfo{author}{\bibfnamefont{A.}~\bibnamefont{Hamed~Moosavian}}
  \bibnamefont{and} \bibinfo{author}{\bibfnamefont{S.}~\bibnamefont{Jordan}},
  \bibinfo{journal}{Phys. Rev. A} \textbf{\bibinfo{volume}{98}},
  \bibinfo{pages}{012332} (\bibinfo{year}{2018}), \eprint{1711.04006}.

\bibitem[{\citenamefont{Moosavian et~al.}(2019)\citenamefont{Moosavian,
  Garrison, and Jordan}}]{Moosavian:2019rxg}
\bibinfo{author}{\bibfnamefont{A.~H.} \bibnamefont{Moosavian}},
  \bibinfo{author}{\bibfnamefont{J.~R.} \bibnamefont{Garrison}},
  \bibnamefont{and} \bibinfo{author}{\bibfnamefont{S.~P.} \bibnamefont{Jordan}}
  (\bibinfo{year}{2019}), \eprint{1911.03505}.

\bibitem[{\citenamefont{Roose et~al.}(2021)\citenamefont{Roose, Bultinck,
  Vanderstraeten, Verstraete, Van~Acoleyen, and Haegeman}}]{roose_lattice_2021}
\bibinfo{author}{\bibfnamefont{G.}~\bibnamefont{Roose}},
  \bibinfo{author}{\bibfnamefont{N.}~\bibnamefont{Bultinck}},
  \bibinfo{author}{\bibfnamefont{L.}~\bibnamefont{Vanderstraeten}},
  \bibinfo{author}{\bibfnamefont{F.}~\bibnamefont{Verstraete}},
  \bibinfo{author}{\bibfnamefont{K.}~\bibnamefont{Van~Acoleyen}},
  \bibnamefont{and} \bibinfo{author}{\bibfnamefont{J.}~\bibnamefont{Haegeman}},
  \bibinfo{journal}{Journal of High Energy Physics}
  \textbf{\bibinfo{volume}{2021}}, \bibinfo{pages}{207} (\bibinfo{year}{2021}),
  ISSN \bibinfo{issn}{1029-8479}, \bibinfo{note}{arXiv: 2010.03441},
  \urlprefix\url{http://arxiv.org/abs/2010.03441}.

\bibitem[{\citenamefont{Roose et~al.}(2022)\citenamefont{Roose, Haegeman,
  Van~Acoleyen, Vanderstraeten, and Bultinck}}]{Roose:2021pba}
\bibinfo{author}{\bibfnamefont{G.}~\bibnamefont{Roose}},
  \bibinfo{author}{\bibfnamefont{J.}~\bibnamefont{Haegeman}},
  \bibinfo{author}{\bibfnamefont{K.}~\bibnamefont{Van~Acoleyen}},
  \bibinfo{author}{\bibfnamefont{L.}~\bibnamefont{Vanderstraeten}},
  \bibnamefont{and} \bibinfo{author}{\bibfnamefont{N.}~\bibnamefont{Bultinck}},
  \bibinfo{journal}{JHEP} \textbf{\bibinfo{volume}{06}}, \bibinfo{pages}{019}
  (\bibinfo{year}{2022}), \eprint{2111.14652}.

\bibitem[{\citenamefont{Jordan and Wigner}(1928)}]{Jordan:1928wi}
\bibinfo{author}{\bibfnamefont{P.}~\bibnamefont{Jordan}} \bibnamefont{and}
  \bibinfo{author}{\bibfnamefont{E.~P.} \bibnamefont{Wigner}},
  \bibinfo{journal}{Z. Phys.} \textbf{\bibinfo{volume}{47}},
  \bibinfo{pages}{631} (\bibinfo{year}{1928}).

\bibitem[{\citenamefont{Dargis and
  Maassarani}(1998)}]{dargis_fermionization_1998}
\bibinfo{author}{\bibfnamefont{P.}~\bibnamefont{Dargis}} \bibnamefont{and}
  \bibinfo{author}{\bibfnamefont{Z.}~\bibnamefont{Maassarani}},
  \bibinfo{journal}{Nuclear Physics B} \textbf{\bibinfo{volume}{535}},
  \bibinfo{pages}{681} (\bibinfo{year}{1998}), ISSN \bibinfo{issn}{05503213},
  \bibinfo{note}{arXiv: cond-mat/9806208},
  \urlprefix\url{http://arxiv.org/abs/cond-mat/9806208}.

\bibitem[{\citenamefont{Reiner et~al.}(2016)\citenamefont{Reiner, Marthaler,
  Braum{\"u}ller, Weides, and Sch{\"o}n}}]{reiner2016emulating}
\bibinfo{author}{\bibfnamefont{J.-M.} \bibnamefont{Reiner}},
  \bibinfo{author}{\bibfnamefont{M.}~\bibnamefont{Marthaler}},
  \bibinfo{author}{\bibfnamefont{J.}~\bibnamefont{Braum{\"u}ller}},
  \bibinfo{author}{\bibfnamefont{M.}~\bibnamefont{Weides}}, \bibnamefont{and}
  \bibinfo{author}{\bibfnamefont{G.}~\bibnamefont{Sch{\"o}n}},
  \bibinfo{journal}{Physical Review A} \textbf{\bibinfo{volume}{94}},
  \bibinfo{pages}{032338} (\bibinfo{year}{2016}).

\bibitem[{\citenamefont{Stanisic et~al.}(2022)\citenamefont{Stanisic, Bosse,
  Gambetta, Santos, Mruczkiewicz, O’Brien, Ostby, and
  Montanaro}}]{stanisic2022observing}
\bibinfo{author}{\bibfnamefont{S.}~\bibnamefont{Stanisic}},
  \bibinfo{author}{\bibfnamefont{J.~L.} \bibnamefont{Bosse}},
  \bibinfo{author}{\bibfnamefont{F.~M.} \bibnamefont{Gambetta}},
  \bibinfo{author}{\bibfnamefont{R.~A.} \bibnamefont{Santos}},
  \bibinfo{author}{\bibfnamefont{W.}~\bibnamefont{Mruczkiewicz}},
  \bibinfo{author}{\bibfnamefont{T.~E.} \bibnamefont{O’Brien}},
  \bibinfo{author}{\bibfnamefont{E.}~\bibnamefont{Ostby}}, \bibnamefont{and}
  \bibinfo{author}{\bibfnamefont{A.}~\bibnamefont{Montanaro}},
  \bibinfo{journal}{Nature communications} \textbf{\bibinfo{volume}{13}},
  \bibinfo{pages}{1} (\bibinfo{year}{2022}).

\bibitem[{\citenamefont{Hubbard}(1963)}]{hubbard1963electron}
\bibinfo{author}{\bibfnamefont{J.}~\bibnamefont{Hubbard}},
  \bibinfo{journal}{Proceedings of the Royal Society of London. Series A.
  Mathematical and Physical Sciences} \textbf{\bibinfo{volume}{276}},
  \bibinfo{pages}{238} (\bibinfo{year}{1963}).

\bibitem[{\citenamefont{Campbell and Bishop}(1982)}]{Campbell:1981dc}
\bibinfo{author}{\bibfnamefont{D.~K.} \bibnamefont{Campbell}} \bibnamefont{and}
  \bibinfo{author}{\bibfnamefont{A.~R.} \bibnamefont{Bishop}},
  \bibinfo{journal}{Nucl. Phys. B} \textbf{\bibinfo{volume}{200}},
  \bibinfo{pages}{297} (\bibinfo{year}{1982}).

\bibitem[{\citenamefont{Chodos and Minakata}(1994)}]{Chodos:1993mf}
\bibinfo{author}{\bibfnamefont{A.}~\bibnamefont{Chodos}} \bibnamefont{and}
  \bibinfo{author}{\bibfnamefont{H.}~\bibnamefont{Minakata}},
  \bibinfo{journal}{Phys. Lett. A} \textbf{\bibinfo{volume}{191}},
  \bibinfo{pages}{39} (\bibinfo{year}{1994}).

\bibitem[{\citenamefont{Kuno}(2019)}]{Kuno:2018pcp}
\bibinfo{author}{\bibfnamefont{Y.}~\bibnamefont{Kuno}}, \bibinfo{journal}{Phys.
  Rev. B} \textbf{\bibinfo{volume}{99}}, \bibinfo{pages}{064105}
  (\bibinfo{year}{2019}), \eprint{1811.01487}.

\bibitem[{\citenamefont{Su et~al.}(1979)\citenamefont{Su, Schrieffer, and
  Heeger}}]{PhysRevLett.42.1698}
\bibinfo{author}{\bibfnamefont{W.~P.} \bibnamefont{Su}},
  \bibinfo{author}{\bibfnamefont{J.~R.} \bibnamefont{Schrieffer}},
  \bibnamefont{and} \bibinfo{author}{\bibfnamefont{A.~J.}
  \bibnamefont{Heeger}}, \bibinfo{journal}{Phys. Rev. Lett.}
  \textbf{\bibinfo{volume}{42}}, \bibinfo{pages}{1698} (\bibinfo{year}{1979}),
  \urlprefix\url{https://link.aps.org/doi/10.1103/PhysRevLett.42.1698}.

\bibitem[{\citenamefont{Basar et~al.}(2009)\citenamefont{Basar, Dunne, and
  Thies}}]{Basar:2009fg}
\bibinfo{author}{\bibfnamefont{G.}~\bibnamefont{Basar}},
  \bibinfo{author}{\bibfnamefont{G.~V.} \bibnamefont{Dunne}}, \bibnamefont{and}
  \bibinfo{author}{\bibfnamefont{M.}~\bibnamefont{Thies}},
  \bibinfo{journal}{Phys. Rev. D} \textbf{\bibinfo{volume}{79}},
  \bibinfo{pages}{105012} (\bibinfo{year}{2009}), \eprint{0903.1868}.

\bibitem[{\citenamefont{Lenz et~al.}(2020)\citenamefont{Lenz, Pannullo, Wagner,
  Wellegehausen, and Wipf}}]{Lenz:2020bxk}
\bibinfo{author}{\bibfnamefont{J.}~\bibnamefont{Lenz}},
  \bibinfo{author}{\bibfnamefont{L.}~\bibnamefont{Pannullo}},
  \bibinfo{author}{\bibfnamefont{M.}~\bibnamefont{Wagner}},
  \bibinfo{author}{\bibfnamefont{B.}~\bibnamefont{Wellegehausen}},
  \bibnamefont{and} \bibinfo{author}{\bibfnamefont{A.}~\bibnamefont{Wipf}},
  \bibinfo{journal}{Phys. Rev. D} \textbf{\bibinfo{volume}{101}},
  \bibinfo{pages}{094512} (\bibinfo{year}{2020}), \eprint{2004.00295}.

\bibitem[{\citenamefont{Bermudez et~al.}(2018)\citenamefont{Bermudez, Tirrito,
  Rizzi, Lewenstein, and Hands}}]{Bermudez:2018eyh}
\bibinfo{author}{\bibfnamefont{A.}~\bibnamefont{Bermudez}},
  \bibinfo{author}{\bibfnamefont{E.}~\bibnamefont{Tirrito}},
  \bibinfo{author}{\bibfnamefont{M.}~\bibnamefont{Rizzi}},
  \bibinfo{author}{\bibfnamefont{M.}~\bibnamefont{Lewenstein}},
  \bibnamefont{and} \bibinfo{author}{\bibfnamefont{S.}~\bibnamefont{Hands}},
  \bibinfo{journal}{Annals Phys.} \textbf{\bibinfo{volume}{399}},
  \bibinfo{pages}{149} (\bibinfo{year}{2018}), \eprint{1807.03202}.

\bibitem[{\citenamefont{Ziegler et~al.}(2020)\citenamefont{Ziegler, Tirrito,
  Lewenstein, Hands, and Bermudez}}]{Ziegler:2020zkq}
\bibinfo{author}{\bibfnamefont{L.}~\bibnamefont{Ziegler}},
  \bibinfo{author}{\bibfnamefont{E.}~\bibnamefont{Tirrito}},
  \bibinfo{author}{\bibfnamefont{M.}~\bibnamefont{Lewenstein}},
  \bibinfo{author}{\bibfnamefont{S.}~\bibnamefont{Hands}}, \bibnamefont{and}
  \bibinfo{author}{\bibfnamefont{A.}~\bibnamefont{Bermudez}}
  (\bibinfo{year}{2020}), \eprint{2011.08744}.

\bibitem[{\citenamefont{Ziegler et~al.}(2022)\citenamefont{Ziegler, Tirrito,
  Lewenstein, Hands, and Bermudez}}]{Ziegler:2021yua}
\bibinfo{author}{\bibfnamefont{L.}~\bibnamefont{Ziegler}},
  \bibinfo{author}{\bibfnamefont{E.}~\bibnamefont{Tirrito}},
  \bibinfo{author}{\bibfnamefont{M.}~\bibnamefont{Lewenstein}},
  \bibinfo{author}{\bibfnamefont{S.}~\bibnamefont{Hands}}, \bibnamefont{and}
  \bibinfo{author}{\bibfnamefont{A.}~\bibnamefont{Bermudez}},
  \bibinfo{journal}{Annals Phys.} \textbf{\bibinfo{volume}{439}},
  \bibinfo{pages}{168763} (\bibinfo{year}{2022}), \eprint{2111.04485}.

\bibitem[{\citenamefont{Tirrito et~al.}(2022)\citenamefont{Tirrito, Lewenstein,
  and Bermudez}}]{Tirrito:2021fbj}
\bibinfo{author}{\bibfnamefont{E.}~\bibnamefont{Tirrito}},
  \bibinfo{author}{\bibfnamefont{M.}~\bibnamefont{Lewenstein}},
  \bibnamefont{and} \bibinfo{author}{\bibfnamefont{A.}~\bibnamefont{Bermudez}},
  \bibinfo{journal}{Phys. Rev. B} \textbf{\bibinfo{volume}{106}},
  \bibinfo{pages}{045147} (\bibinfo{year}{2022}), \eprint{2112.07654}.

\bibitem[{\citenamefont{Surace et~al.}(2020)\citenamefont{Surace, Mazza,
  Giudici, Lerose, Gambassi, and Dalmonte}}]{Surace:2019dtp}
\bibinfo{author}{\bibfnamefont{F.~M.} \bibnamefont{Surace}},
  \bibinfo{author}{\bibfnamefont{P.~P.} \bibnamefont{Mazza}},
  \bibinfo{author}{\bibfnamefont{G.}~\bibnamefont{Giudici}},
  \bibinfo{author}{\bibfnamefont{A.}~\bibnamefont{Lerose}},
  \bibinfo{author}{\bibfnamefont{A.}~\bibnamefont{Gambassi}}, \bibnamefont{and}
  \bibinfo{author}{\bibfnamefont{M.}~\bibnamefont{Dalmonte}},
  \bibinfo{journal}{Phys. Rev. X} \textbf{\bibinfo{volume}{10}},
  \bibinfo{pages}{021041} (\bibinfo{year}{2020}), \eprint{1902.09551}.

\bibitem[{\citenamefont{Surace and Lerose}(2021)}]{Surace:2020ycc}
\bibinfo{author}{\bibfnamefont{F.~M.} \bibnamefont{Surace}} \bibnamefont{and}
  \bibinfo{author}{\bibfnamefont{A.}~\bibnamefont{Lerose}},
  \bibinfo{journal}{New J. Phys.} \textbf{\bibinfo{volume}{23}},
  \bibinfo{pages}{062001} (\bibinfo{year}{2021}), \eprint{2011.10583}.

\bibitem[{\citenamefont{Notarnicola et~al.}(2020)\citenamefont{Notarnicola,
  Collura, and Montangero}}]{Notarnicola:2019wzb}
\bibinfo{author}{\bibfnamefont{S.}~\bibnamefont{Notarnicola}},
  \bibinfo{author}{\bibfnamefont{M.}~\bibnamefont{Collura}}, \bibnamefont{and}
  \bibinfo{author}{\bibfnamefont{S.}~\bibnamefont{Montangero}},
  \bibinfo{journal}{Phys. Rev. Res.} \textbf{\bibinfo{volume}{2}},
  \bibinfo{pages}{013288} (\bibinfo{year}{2020}), \eprint{1907.12579}.

\bibitem[{\citenamefont{Meurice}(2021{\natexlab{a}})}]{Meurice:2021pvj}
\bibinfo{author}{\bibfnamefont{Y.}~\bibnamefont{Meurice}},
  \bibinfo{journal}{Phys. Rev. D} \textbf{\bibinfo{volume}{104}},
  \bibinfo{pages}{094513} (\bibinfo{year}{2021}{\natexlab{a}}),
  \eprint{2107.11366}.

\bibitem[{\citenamefont{Verschelde et~al.}(1997)\citenamefont{Verschelde,
  Schelstraete, and Vanderkelen}}]{Verschelde:1997jx}
\bibinfo{author}{\bibfnamefont{H.}~\bibnamefont{Verschelde}},
  \bibinfo{author}{\bibfnamefont{S.}~\bibnamefont{Schelstraete}},
  \bibnamefont{and}
  \bibinfo{author}{\bibfnamefont{M.}~\bibnamefont{Vanderkelen}},
  \bibinfo{journal}{Z. Phys. C} \textbf{\bibinfo{volume}{76}},
  \bibinfo{pages}{161} (\bibinfo{year}{1997}).

\bibitem[{\citenamefont{Choi et~al.}(2017)\citenamefont{Choi, Ryttov, and
  Shrock}}]{Choi:2016sxt}
\bibinfo{author}{\bibfnamefont{G.}~\bibnamefont{Choi}},
  \bibinfo{author}{\bibfnamefont{T.~A.} \bibnamefont{Ryttov}},
  \bibnamefont{and} \bibinfo{author}{\bibfnamefont{R.}~\bibnamefont{Shrock}},
  \bibinfo{journal}{Phys. Rev. D} \textbf{\bibinfo{volume}{95}},
  \bibinfo{pages}{025012} (\bibinfo{year}{2017}), \eprint{1612.05580}.

\bibitem[{\citenamefont{Trotter}(1959)}]{trotter1959product}
\bibinfo{author}{\bibfnamefont{H.~F.} \bibnamefont{Trotter}},
  \bibinfo{journal}{Proceedings of the American Mathematical Society}
  \textbf{\bibinfo{volume}{10}}, \bibinfo{pages}{545} (\bibinfo{year}{1959}).

\bibitem[{\citenamefont{Suzuki}(1992)}]{suzuki1992general}
\bibinfo{author}{\bibfnamefont{M.}~\bibnamefont{Suzuki}},
  \bibinfo{journal}{Physics Letters A} \textbf{\bibinfo{volume}{165}},
  \bibinfo{pages}{387} (\bibinfo{year}{1992}).

\bibitem[{\citenamefont{Suzuki}(1990)}]{suzuki1990fractal}
\bibinfo{author}{\bibfnamefont{M.}~\bibnamefont{Suzuki}},
  \bibinfo{journal}{Physics Letters A} \textbf{\bibinfo{volume}{146}},
  \bibinfo{pages}{319} (\bibinfo{year}{1990}).

\bibitem[{\citenamefont{Suzuki}(1993)}]{suzuki1993improved}
\bibinfo{author}{\bibfnamefont{M.}~\bibnamefont{Suzuki}},
  \bibinfo{journal}{Physics Letters A} \textbf{\bibinfo{volume}{180}},
  \bibinfo{pages}{232} (\bibinfo{year}{1993}).

\bibitem[{\citenamefont{Melzer}(1995)}]{melzer_scaling_1995}
\bibinfo{author}{\bibfnamefont{E.}~\bibnamefont{Melzer}},
  \bibinfo{journal}{Nuclear Physics B} \textbf{\bibinfo{volume}{443}},
  \bibinfo{pages}{553} (\bibinfo{year}{1995}), ISSN \bibinfo{issn}{05503213},
  \bibinfo{note}{arXiv: cond-mat/9410043},
  \urlprefix\url{http://arxiv.org/abs/cond-mat/9410043}.

\bibitem[{\citenamefont{Ayyar and Chandrasekharan}(2015)}]{Ayyar:2014eua}
\bibinfo{author}{\bibfnamefont{V.}~\bibnamefont{Ayyar}} \bibnamefont{and}
  \bibinfo{author}{\bibfnamefont{S.}~\bibnamefont{Chandrasekharan}},
  \bibinfo{journal}{Phys. Rev. D} \textbf{\bibinfo{volume}{91}},
  \bibinfo{pages}{065035} (\bibinfo{year}{2015}), \eprint{1410.6474}.

\bibitem[{\citenamefont{Catterall}(2016)}]{Catterall:2015zua}
\bibinfo{author}{\bibfnamefont{S.}~\bibnamefont{Catterall}},
  \bibinfo{journal}{JHEP} \textbf{\bibinfo{volume}{01}}, \bibinfo{pages}{121}
  (\bibinfo{year}{2016}), \eprint{1510.04153}.

\bibitem[{\citenamefont{Ayyar and Chandrasekharan}(2017)}]{Ayyar:2017qii}
\bibinfo{author}{\bibfnamefont{V.}~\bibnamefont{Ayyar}} \bibnamefont{and}
  \bibinfo{author}{\bibfnamefont{S.}~\bibnamefont{Chandrasekharan}},
  \bibinfo{journal}{Phys. Rev. D} \textbf{\bibinfo{volume}{96}},
  \bibinfo{pages}{114506} (\bibinfo{year}{2017}), \eprint{1709.06048}.

\bibitem[{\citenamefont{Ayyar and Chandrasekharan}(2016)}]{Ayyar:2015lrd}
\bibinfo{author}{\bibfnamefont{V.}~\bibnamefont{Ayyar}} \bibnamefont{and}
  \bibinfo{author}{\bibfnamefont{S.}~\bibnamefont{Chandrasekharan}},
  \bibinfo{journal}{Phys. Rev. D} \textbf{\bibinfo{volume}{93}},
  \bibinfo{pages}{081701} (\bibinfo{year}{2016}), \eprint{1511.09071}.

\bibitem[{\citenamefont{Butt et~al.}(2018)\citenamefont{Butt, Catterall, and
  Schaich}}]{Butt:2018nkn}
\bibinfo{author}{\bibfnamefont{N.}~\bibnamefont{Butt}},
  \bibinfo{author}{\bibfnamefont{S.}~\bibnamefont{Catterall}},
  \bibnamefont{and} \bibinfo{author}{\bibfnamefont{D.}~\bibnamefont{Schaich}},
  \bibinfo{journal}{Phys. Rev. D} \textbf{\bibinfo{volume}{98}},
  \bibinfo{pages}{114514} (\bibinfo{year}{2018}), \eprint{1810.06117}.

\bibitem[{\citenamefont{Meurice}(2021{\natexlab{b}})}]{meurice2021quantum}
\bibinfo{author}{\bibfnamefont{Y.}~\bibnamefont{Meurice}},
  \emph{\bibinfo{title}{Quantum Field Theory}} (\bibinfo{publisher}{IOP
  Publishing}, \bibinfo{year}{2021}{\natexlab{b}}).

\bibitem[{\citenamefont{Fishman et~al.}(2020)\citenamefont{Fishman, White, and
  Stoudenmire}}]{itensor}
\bibinfo{author}{\bibfnamefont{M.}~\bibnamefont{Fishman}},
  \bibinfo{author}{\bibfnamefont{S.~R.} \bibnamefont{White}}, \bibnamefont{and}
  \bibinfo{author}{\bibfnamefont{E.~M.} \bibnamefont{Stoudenmire}},
  \emph{\bibinfo{title}{The \mbox{ITensor} software library for tensor network
  calculations}} (\bibinfo{year}{2020}), \eprint{2007.14822}.

\bibitem[{\citenamefont{McClean et~al.}(2016)\citenamefont{McClean, Romero,
  Babbush, and Aspuru-Guzik}}]{mcclean2016theory}
\bibinfo{author}{\bibfnamefont{J.~R.} \bibnamefont{McClean}},
  \bibinfo{author}{\bibfnamefont{J.}~\bibnamefont{Romero}},
  \bibinfo{author}{\bibfnamefont{R.}~\bibnamefont{Babbush}}, \bibnamefont{and}
  \bibinfo{author}{\bibfnamefont{A.}~\bibnamefont{Aspuru-Guzik}},
  \bibinfo{journal}{New Journal of Physics} \textbf{\bibinfo{volume}{18}},
  \bibinfo{pages}{023023} (\bibinfo{year}{2016}).

\bibitem[{\citenamefont{Kandala et~al.}(2017)\citenamefont{Kandala, Mezzacapo,
  Temme, Takita, Brink, Chow, and Gambetta}}]{kandala2017hardware}
\bibinfo{author}{\bibfnamefont{A.}~\bibnamefont{Kandala}},
  \bibinfo{author}{\bibfnamefont{A.}~\bibnamefont{Mezzacapo}},
  \bibinfo{author}{\bibfnamefont{K.}~\bibnamefont{Temme}},
  \bibinfo{author}{\bibfnamefont{M.}~\bibnamefont{Takita}},
  \bibinfo{author}{\bibfnamefont{M.}~\bibnamefont{Brink}},
  \bibinfo{author}{\bibfnamefont{J.~M.} \bibnamefont{Chow}}, \bibnamefont{and}
  \bibinfo{author}{\bibfnamefont{J.~M.} \bibnamefont{Gambetta}},
  \bibinfo{journal}{Nature} \textbf{\bibinfo{volume}{549}},
  \bibinfo{pages}{242} (\bibinfo{year}{2017}).

\bibitem[{\citenamefont{MacDonald}(1933)}]{macdonald1933successive}
\bibinfo{author}{\bibfnamefont{J.}~\bibnamefont{MacDonald}},
  \bibinfo{journal}{Physical Review} \textbf{\bibinfo{volume}{43}},
  \bibinfo{pages}{830} (\bibinfo{year}{1933}).

\bibitem[{\citenamefont{Powell}(1994)}]{powell1994direct}
\bibinfo{author}{\bibfnamefont{M.~J.} \bibnamefont{Powell}}, in
  \emph{\bibinfo{booktitle}{Advances in optimization and numerical analysis}}
  (\bibinfo{publisher}{Springer}, \bibinfo{year}{1994}), pp.
  \bibinfo{pages}{51--67}.

\bibitem[{\citenamefont{Powell}(1998)}]{powell1998direct}
\bibinfo{author}{\bibfnamefont{M.~J.} \bibnamefont{Powell}},
  \bibinfo{journal}{Acta numerica} \textbf{\bibinfo{volume}{7}},
  \bibinfo{pages}{287} (\bibinfo{year}{1998}).

\bibitem[{\citenamefont{Powell}(2007)}]{powell2007view}
\bibinfo{author}{\bibfnamefont{M.~J.} \bibnamefont{Powell}},
  \bibinfo{journal}{Mathematics Today-Bulletin of the Institute of Mathematics
  and its Applications} \textbf{\bibinfo{volume}{43}}, \bibinfo{pages}{170}
  (\bibinfo{year}{2007}).

\bibitem[{\citenamefont{Kraft}(1988)}]{kraft1988software}
\bibinfo{author}{\bibfnamefont{D.}~\bibnamefont{Kraft}},
  \bibinfo{journal}{Forschungsbericht- Deutsche Forschungs- und Versuchsanstalt
  fur Luft- und Raumfahrt}  (\bibinfo{year}{1988}).

\bibitem[{\citenamefont{Ferris}(2014)}]{ferris2014fourier}
\bibinfo{author}{\bibfnamefont{A.~J.} \bibnamefont{Ferris}},
  \bibinfo{journal}{Phys. Rev. Lett.} \textbf{\bibinfo{volume}{113}},
  \bibinfo{pages}{010401} (\bibinfo{year}{2014}),
  \urlprefix\url{https://link.aps.org/doi/10.1103/PhysRevLett.113.010401}.

\bibitem[{\citenamefont{Kivlichan et~al.}(2020)\citenamefont{Kivlichan, Gidney,
  Berry, Wiebe, McClean, Sun, Jiang, Rubin, Fowler, Aspuru-Guzik
  et~al.}}]{kivlichan2020improved}
\bibinfo{author}{\bibfnamefont{I.~D.} \bibnamefont{Kivlichan}},
  \bibinfo{author}{\bibfnamefont{C.}~\bibnamefont{Gidney}},
  \bibinfo{author}{\bibfnamefont{D.~W.} \bibnamefont{Berry}},
  \bibinfo{author}{\bibfnamefont{N.}~\bibnamefont{Wiebe}},
  \bibinfo{author}{\bibfnamefont{J.}~\bibnamefont{McClean}},
  \bibinfo{author}{\bibfnamefont{W.}~\bibnamefont{Sun}},
  \bibinfo{author}{\bibfnamefont{Z.}~\bibnamefont{Jiang}},
  \bibinfo{author}{\bibfnamefont{N.}~\bibnamefont{Rubin}},
  \bibinfo{author}{\bibfnamefont{A.}~\bibnamefont{Fowler}},
  \bibinfo{author}{\bibfnamefont{A.}~\bibnamefont{Aspuru-Guzik}},
  \bibnamefont{et~al.}, \bibinfo{journal}{Quantum}
  \textbf{\bibinfo{volume}{4}}, \bibinfo{pages}{296} (\bibinfo{year}{2020}).

\bibitem[{\citenamefont{White}(1992)}]{PhysRevLett.69.2863}
\bibinfo{author}{\bibfnamefont{S.~R.} \bibnamefont{White}},
  \bibinfo{journal}{Phys. Rev. Lett.} \textbf{\bibinfo{volume}{69}},
  \bibinfo{pages}{2863} (\bibinfo{year}{1992}),
  \urlprefix\url{https://link.aps.org/doi/10.1103/PhysRevLett.69.2863}.

\bibitem[{\citenamefont{White}(1993)}]{PhysRevB.48.10345}
\bibinfo{author}{\bibfnamefont{S.~R.} \bibnamefont{White}},
  \bibinfo{journal}{Phys. Rev. B} \textbf{\bibinfo{volume}{48}},
  \bibinfo{pages}{10345} (\bibinfo{year}{1993}),
  \urlprefix\url{https://link.aps.org/doi/10.1103/PhysRevB.48.10345}.

\bibitem[{\citenamefont{Schollw\"ock}(2005)}]{RevModPhys.77.259}
\bibinfo{author}{\bibfnamefont{U.}~\bibnamefont{Schollw\"ock}},
  \bibinfo{journal}{Rev. Mod. Phys.} \textbf{\bibinfo{volume}{77}},
  \bibinfo{pages}{259} (\bibinfo{year}{2005}),
  \urlprefix\url{https://link.aps.org/doi/10.1103/RevModPhys.77.259}.

\end{thebibliography}

\end{document}